\documentclass[preprint,12pt]{elsarticle}

\usepackage{natbib}

\usepackage{algorithm}
\usepackage{algpseudocode}
\usepackage{eurosym}
\usepackage{float}

\usepackage{amsmath}

\usepackage{amssymb}

\usepackage{graphicx}
\usepackage[T1]{fontenc}
\usepackage[utf8]{inputenc}
\usepackage{flushend}
\usepackage{chngcntr}
\usepackage{url}
\usepackage{listings}
\usepackage{fancyhdr}
\usepackage[normalem]{ulem}
\usepackage{subfig}

\usepackage{enumerate}
\usepackage[shortlabels]{enumitem}

\usepackage{xcolor}

\usepackage{enumerate}
\usepackage{tcolorbox}

\usepackage{booktabs}
\usepackage{pifont}
\usepackage{lscape}
\usepackage{rotating}
\usepackage{bibentry}

\usepackage{pdfpages}
\usepackage{adjustbox}

\usepackage{ccommands}

\usepackage{multirow}
\usepackage{xcolor}

\newcommand{\orcidX}[1]{\href{https://orcid.org/#1}{\includegraphics[width=10pt]{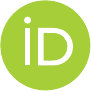}}}

\usepackage{amssymb}
\usepackage{amsmath}

\journal{–}

\begin{document}

\begin{frontmatter}



\title{An Empirical Study on the Classification of Bug Reports with Machine Learning}


\author[UC]{Renato Andrade}
\ead{renatoandrade@dei.uc.pt}
\author[UC]{César Teixeira}
\ead{cteixei@dei.uc.pt}
\author[UC]{Nuno Laranjeiro}
\ead{cnl@dei.uc.pt}
\author[UNCC]{Marco Vieira}
\ead{marco.vieira@charlotte.edu}

\affiliation[UC]{organization={University of Coimbra, CISUC/LASI, DEI},
            country={Portugal}}

\affiliation[UNCC]{organization={University of North Carolina at Charlotte},
            country={USA}}


\begin{abstract}

Software defects are a major threat to the reliability of computer systems. The literature shows that more than 30\% of bug reports submitted in large software projects are misclassified (i.e., are feature requests, or mistakes made by the bug reporter), leading developers to place great effort in manually inspecting them. Machine Learning algorithms can be used for the automatic classification of issue reports. Still, little is known regarding key aspects of training models, such as the influence of programming languages and issue tracking systems. In this paper, we use a dataset containing more than 660,000 issue reports, collected from heterogeneous projects hosted in different issue tracking systems, to study how different factors (e.g., project language, report content) can influence the performance of models in handling classification of issue reports. Results show that using the report title or description does not significantly differ; Support Vector Machine, Logistic Regression, and Random Forest are effective in classifying issue reports; programming languages and issue tracking systems influence classification outcomes; and models based on heterogeneous projects can classify reports from projects not present during training. Based on findings, we propose guidelines for future research, including recommendations for using heterogeneous data and selecting high-performing algorithms.

\end{abstract}




\end{frontmatter}

\newcommand{\cmark}{\ding{51}}%
\newcommand{\xmark}{\ding{55}}%

\section{Introduction} 
\label{sec_introduction}

A bug-fixing process should be triggered whenever a software fault is discovered (e.g., by a user) in a system, to allow developers to resolve the issue. For instance, in open-source communities (e.g., \texttt{apache.org}), users and developers typically create issue reports that are submitted via an issue tracking system (ITS) \cite{Zhu2019}, such as BugZilla, GitHub or Jira. 
Issue reports are semi-structured documents that report situations or problems faced when using or testing a software system. They usually include attributes that provide information about the context of use and plain text fields, namely a title and a description. These reports are then analyzed by developers concerning a few fundamental aspects, including the severity of the issue, the developer that should be responsible for handling it, and, especially, if the report is correctly labelled or not (i.e., whether it is indeed the type of issue that its creator says it is).

Previous work has shown that a large number of the issues reported as bugs in popular open source systems are actually not bugs, being instead requests for new features, improvements, documentation updates, or simply mistakes by the issue reporters \cite{Herzig2013,Lopes2020,Catolino2019,Terdchanakul2017}. This is frequently due to the lack of technical knowledge about the system, as many issue reporters are end users \cite{Chawla2015}. The problem is that the manual analysis of a single issue is complex and time consuming (potentially involving several developers) and frequently results in wasted time and resources, when developers conclude that a bug report actually does not refer to a bug at all.

Machine Learning (ML) techniques are nowadays increasingly being used for automating various tasks, including manual processes in software engineering, such as bug fixing-related procedures and other associated tasks (e.g., bug triage, defect type classification) \cite{Lopes2020,Kukkar2019,Liu2022,ArcelliFontana2017,Azeem2019}. However, in what concerns the apparently simple task of deciding whether a reported issue is a bug or not, there are several difficulties, which arise from the complexity of dealing with unstructured data and from the vast space available for various configurations and selections, such as the ones referring to feature engineering and model tuning (e.g., hyperparameter selection). Although work has been carried out on this topic (e.g., \cite{Catolino2019,Kochhar2014,Qin2018}), there is still scarce information regarding which data should be used to create models and what is their real effectiveness, especially when considering cases where the reports being classified are intrinsically different from the ones used to train the algorithms.

In this empirical work, we study the use of Machine Learning algorithms (Support Vector Machines, Random Forest, Logistic Regression, Na{\"i}ve Bayes and K-Nearest Neighbors) to classify bug reports, considering the main purpose of distinguishing if a particular issue refers to a real bug or not, in various scenarios. Our option was set on classical machine learning algorithms not only due to their need for less data and computational resources, but mostly due to the unclear and often contradictory results present in the literature for this class of algorithms \cite{Zhou2016, ko2006, Limsettho2016}. For this, we created a dataset with more than 660,000 issue report samples,  of which we use \textit{fixed} and \textit{resolved} reports (i.e., bugs that were acknowledged by developers and for which a fix was made available and is already integrated into the project). These reports were collected from 52 software systems, written in multiple programming languages (e.g., Java, Python, C/C++, JavaScript), and managed in three distinct ITSs (Jira, GitHub and BugZilla). We present a set of experiments and analyze the overall effectiveness of the ML algorithms considering different aspects, such as the best suitable type of information to train the algorithms, how well different algorithms handle the automatic classification of issue reports, the influence of programming languages and issue tracking systems in the performance of the models, and how well do classifiers generalize when facing new reports (e.g., from other projects not present in the training phase). 

Our results show that ML algorithms can effectively classify bug reports, although performance largely depends on the specific scenario being considered. As main observations, we highlight that using either the title or the complete description of bug reports during the training phase does not lead to statistically significant differences in the performance of the models; Support Vector Machine, Logistic Regression and Random Forest are associated with better results in comparison with other algorithms; and programming languages and issue tracking systems may influence the performance of the models. We also observe no significant differences in most cases regarding models trained and tested with reports from a specific system \textit{versus} models tested with reports from projects not used during training, as long as the programming language and ITS are the same. Based on these results, we identify key aspects related with the machine learning pipeline being used in this research topic, such as criteria to consider in data acquisition (e.g., heterogeneous data), or the selection of feature engineering techniques. 

The main contributions of this paper are the following:

\begin{itemize}


    \item An empirical analysis of the overall performance of classical machine learning models in the binary classification of issue reports, based on an heterogeneous labeled dataset of bug reports extracted from three well-known ITSs (GitHub, BugZilla and Jira), referring to 52 open-source software systems, including well-known projects like Elasticsearch, Apache Cassandra and Mozilla Firefox, written in various programming languages like Java, JavaScript, C/C++, and Python;

    \item The analyis of the effect of critical properties of the training process and data, including the model configuration and the content of the report that should be used, influence of the programming language, impact of the tracking system, and ability to generalize when facing new reports (i.e., cross-project classification).


    \item The identification of guidelines to support future research on the automatic classification of issue reports using machine learning, including configuration to follow within the machine learning pipeline (e.g., criteria for data acquisition, feature engineering options).
        
\end{itemize}

   
The rest of this paper is organized as follows. Section \ref{sec_related} presents related work on methodologies for bug report classification using ML. Section \ref{sec_study} defines the research questions and approach. Sections \ref{section_res_rq1} to \ref{section_res_rq5} discuss the results obtained for different research questions. Section \ref{sec_guidelines} compiles a set of guidelines for future research on the automatic classification of bug reports. Section \ref{sec_thereats} discusses the threats to validity and, finally, Section \ref{sec_conclusion} concludes the paper.
\section{Related Work} 
\label{sec_related}
Several approaches can be found in the literature for the automatic classification of issue reports using Machine Learning. Such works generally fit in four main groups: binary classification of bug reports (i.e., bug \textit{versus} non-bug) \cite{Sohrawardi2014,Qin2018}; severity classification \cite{Shatnawi2022,Pushpalatha2019}; triaging (i.e., developer handover) \cite{Zhang2020,Liu2022}; and duplicate report detection \cite{He2020,Kukkar2020}. Across all of these, we find various techniques being used, including deep learning \cite{Lee2017,Mian2021} and classical machine learning techniques (i.e., learning algorithms that do not rely on neural networks, such as decision trees, support vector machines, k-nearest neighbors, or linear regression) \cite{Fan2017,Kukkar2018}. In this paper the focus is set on classical machine learning, due to the  the diverse and often contradictory classification results identified in the literature for this class of algorithms \cite{Zhou2016, ko2006, Limsettho2016}. 
It is worthwhile mentioning that while we discuss performance metrics, the goal is larger than performance, aiming at understanding how various factors in the context of the bug reports may influence performance (e.g., programming language, report content).

Table \ref{tab_related_work} overviews previous research on bug report classification that uses classical machine learning algorithms, considering key aspects 
such as the number of examples used in each study, as well as their origin (i.e., Jira, BugZilla or Github). The table also details which preprocessing techniques were used in each case, whether the approach uses the Bag-of-Words model (BoW) or not, if titles or descriptions are considered, and if the authors opted for using extra attributes (e.g., structured attributes) together with features extracted from plain text. Details such as the classifiers used in the studies, the number of classes, whether the work uses cross-validation, split-validation, or both, and the results obtained are also shown. In the results column, accuracy is abbreviated to \textit{A}, precision to \textit{P}, recall to \textit{R}, F-measure to \textit{F1}, error rate to \textit{ER} and cluster purity to \textit{CP}.

Previous approaches usually rely on features extracted from issue report titles, descriptions or comments, which are textual fields, instead of using other sources of information that may not be present in different issue tracking systems \cite{Kallis2019,Pingclasai2013,Limsettho2014,Limsettho2016}. Thus, different methods to generate features using plain text from issue reports are considered, although most authors use Bag-of-Words (BoW), which is a standard representation for textual information \cite{Srinivasa2018}. Yet, currently known approaches are often based on different classifiers and use different evaluation metrics \cite{Zhu2019,Pandey2017}.

\begin{table}[h!]
\caption{Comparison of state-of-art approaches for automatic classification of issue reports.}\label{tab_related_work}
\centering
\includegraphics[width=0.78\textwidth]{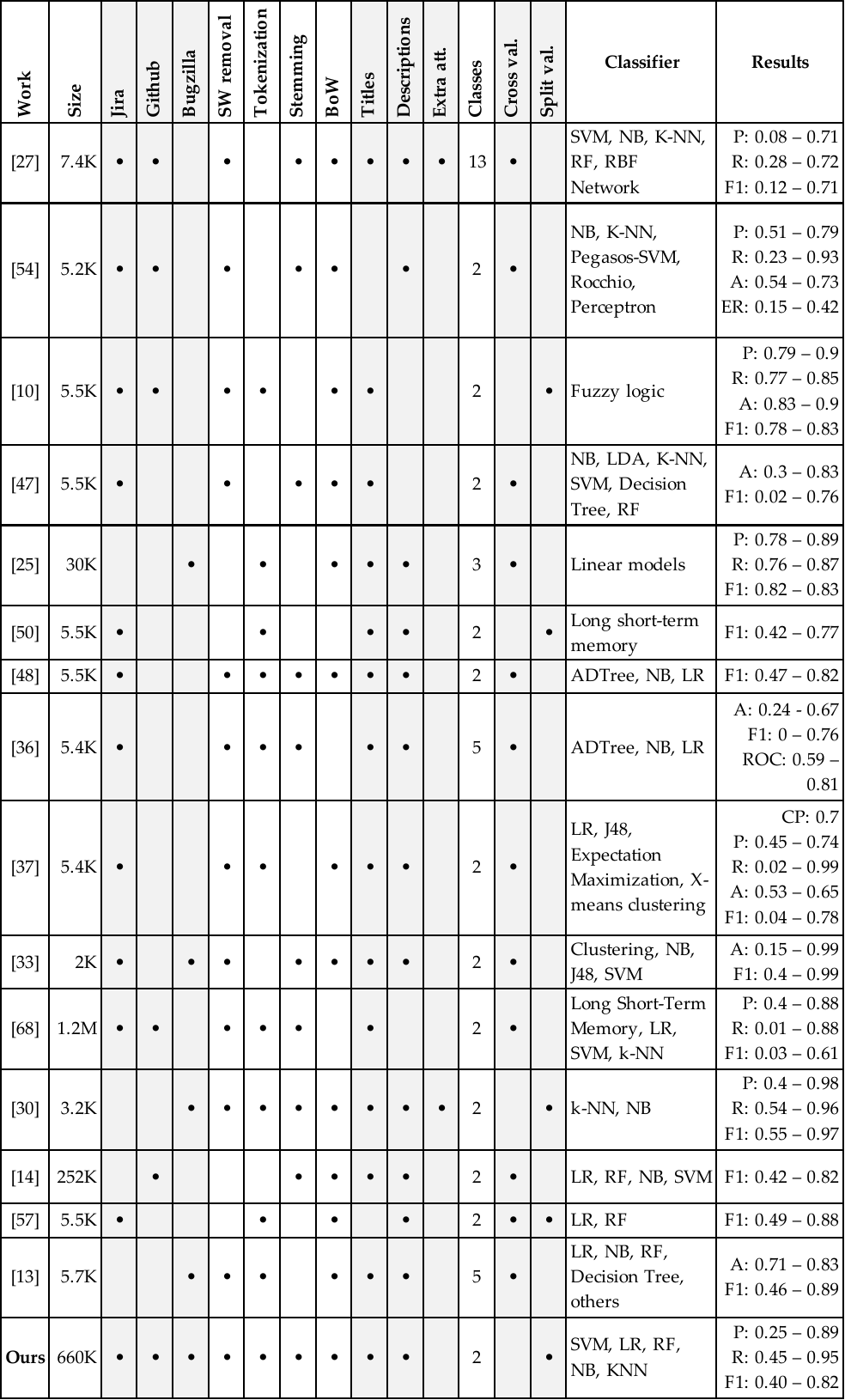}

\end{table}

We observe key limitations in previous works, as follows. The first is related with the \textbf{dimension of the data}. Several authors mention that their datasets have a given number of issue reports; however, to train the models, they only use a small percentage of their samples, which in some cases is less than 15\% of the dataset.\cite{Pandey2017,Sohrawardi2014} 
This usually occurs as they are using data from various projects, so when isolated, none of them has a reasonable number of issue reports to be used.

Another critical aspect noticeable in current studies is \textbf{the use of the same datasets} in different works. As can be noticed in Table \ref{tab_related_work}, besides using relatively small datasets (e.g., 2,000 to 7,400 issue reports), most studies use the same data (such as the dataset created in \cite{Herzig2013}) as issue reports correctly labeled can be difficult to compile 
\cite{Chawla2015,Qin2018,Pingclasai2013,Pandey2017}. This limits the potential of current approaches, making them difficult to generalize, considering that many of them use projects written in the same programming language, usually extracted from the same ITS.

The \textbf{content of the reports used to train the models} is also an aspect to consider. Many authors suggest that the use of titles is sufficient to train models for the classification of issue reports (e.g., \cite{Zhou2016, ko2006}), while others argue that titles are too short and do not provide the information needed for such a purpose (e.g., \cite{Limsettho2016}).
Regarding the \textbf{use of structured attributes} from the issue reports, we notice that some authors report that they achieved high levels of accuracy or F-measure in their experiments (e.g., \cite{Kochhar2014,Kukkar2018}) when using structured data to create their models. The problem is that, many ITSs like GitHub for example, do not provide such information, making it difficult to reproduce their work.

Equally concerning is the \textbf{use of inadequate attributes}. The literature reveals that some studies resort to questionable data to train their models \cite{Kochhar2014}. A clear example is the adoption of an attribute called \textit{reporter}, which is the person who reported the issue. In some situations this may makes sense, for example, when a specific group of people is responsible for creating the reports. However, in open-source projects, which are used in most studies, anyone can report an issue, making the reporter an inappropriate attribute to be used, as the models will struggle to classify issues reported by new users whenever such attribute has a substantial weight.

Regarding the \textbf{choice of classifiers}, many works (e.g., \cite{Pingclasai2013,Limsettho2014} and \cite{Fan2017,Du2017}) are based on the same set of classifiers, some of them not really appropriate for the context. For instance, algorithms like a simple decision tree can show promising performance in small datasets, where they easily become very adapted to the data \cite{Pandey2017}. However, depending on the features and the pruning strategy, a classical decision tree might not be the most suitable choice \cite{MingersBsrcd1989}, as they may struggle to classify samples that do not contain specific terms in some of their nodes, even when the issue reports have other relevant words instead.

One aspect we frequently observe is that \textbf{empirical evidence is insufficient} in many works, with authors sometimes claiming that a given method has better results than others although not accompanying such claims with the corresponding data (e.g., \cite{Chawla2015}). The great majority of studies in this context,  including \cite{Kochhar2014,KumarNagwani2012,Limsettho2014,Limsettho2016}, does not apply formal hypothesis testing procedures to analyse results.

In this paper, we use data collected from heterogeneous projects which are supported by different tracking systems. We train models with both titles and descriptions, to analyze which report content leads to better results. We focus on plain text interpretation for feature generation, as some ITSs do not provide structured attributes. As classifiers, we use five classical machine learning algorithms, to understand the overall effectiveness in the same conditions, as well as the effect of potentially influencing aspects in the classification effectiveness (e.g., programming language of the bug being reported, content of the report). As mentioned, the reason for opting for classical machine learning algorithms has to due with the high heterogeneity of classification effectiveness results reported in related work.  

\section{Study Design} 
\label{sec_study}

This section presents our research questions and describes the approach followed, from the perspective of a typical Machine Learning pipeline \cite{hapke2020,Sammut2011}. In practice, the study aims at contributing to answering the following Research Questions (RQs):

\begin{itemize}
    \item \textbf{RQ1:} Which bug report information (i.e., report title \textit{vs} detailed description) is more useful for training ML models to perform automatic bug report classification?
    \item \textbf{RQ2}: Which ML algorithms achieve better performance in the automatic classification of issue reports? 
    \item \textbf{RQ3}: Does the (main) programming language in which the code is written influence the effectiveness of bug reports classification?
    \item \textbf{RQ4}: Does the tracking system used in the software project influence the effectiveness of bug reports classification?
    \item \textbf{RQ5}: How well do ML models generalize when facing new reports (i.e., from projects not present in the training phase)?
\end{itemize}

To answer \textbf{RQ1}, we start by analyzing which type of unstructured information is more valuable to classify issue reports. We decided to address this aspect because previous studies (e.g., \cite{Zhou2016, ko2006}) suggest that report titles are enough to train models in this context (as opposed to using the full textual description of the issue reports), while some other authors (e.g.,  \cite{Limsettho2016}), argue that titles are too short and do not contain enough information, but precise evidence is missing in the literature. 

Also in the scope of our first RQ, we investigate the best suitable number of dimensions to use. When using a Bag-of-Words (BoW) approach, which is our case, the total amount of features or dimensions is equivalent to the vocabulary size. This can lead to a problem typically known as \textit{curse of dimensionality} \cite{Aggarwal2015,Verleysen2005}. To avoid this problem, we analyze what is a suitable number of dimensions to train models for the classification of issue reports. Using smaller amounts of data is generally a better option (due to computational and resource usage reasons), but there is no evidence of what the minimum is when it comes to the number of dimensions. In practice, an increasing number of dimensions requires more samples to extract enough values for the features, increasing the chances of overfitting.

Concerning \textbf{RQ2}, our objective is to investigate which classifier presents the overall best performance when handling the automatic classification of issue reports. Although some authors addressed this issue in the past (e.g., \cite{Kochhar2014,Fan2017}), the methodologies followed often use quite small datasets, composed of data that is extracted from projects that are very similar in nature (i.e., projects written in the same programming language and that use the same tracking system), which leads to results that do not really represent the classification effectiveness of the algorithms used.

As mentioned before, we use a dataset containing more than 660,000 issue reports, which is 
120 times larger than the median size of the datasets that we found in related work. We pay special attention to the inclusion of heterogeneous data, multiple projects of diverse scales, software developed in different programming languages, and projects using different issue tracking systems and supported by various communities, with the goal of understanding the overall performance of ML algorithms in a representative manner. We consider five well-known algorithms: k-Nearest Neighbor classifier (k-NN) \cite{Cunningham2021}, Naïve Bayes (NB) \cite{Kotu2019}, the Support Vector Machine (SVM) \cite{Cortes1992}, Random Forest (RF) \cite{Breiman2001}, and Logistic Regression (LR), which are known to have been used in related work and for which heterogenous results are known (refer to Table \ref{tab_related_work} for an overview). These are further discussed in Section \ref{subsec_model_selection}. 

Regarding \textbf{RQ3}, the goal is to understand if the main programming language of the project to which the issue report is associated contributes (or not) to significant differences in the classification effectiveness. Notice that nowadays it is very usual for projects to be written using multiple programming languages (e.g., a project may use backend services written in Python along with a frontend fully developed in Javascript). Thus, we refer to the \textit{main} programming language of the project as being the one in which the majority of the code is written. 
The motivation for pursuing this questions is that our analysis of the literature does not allow concluding if the programming language, which for instance, may be associated with distinct issue report writing patterns by developers, is a relevant factor in the classification effectiveness or not \cite{Sohrawardi2014,Kallis2019,Zhu2019,Fan2017}.

In \textbf{RQ4}, we focus our attention in the system used to report the bugs. The hypothesis is that the performance of classifiers might be influenced by the way different issue tracking systems (ITSs) allow developers to report bugs (e.g., by limiting the size of the fields used to report the bug, or by forcing the reporter to place information in certain predefined fields). While several authors analyzed GitHub issue reports (e.g., \cite{Kallis2019, Fan2017}), others used Jira reports (e.g., \cite{Pandey2017, Limsettho2016}) and others opted by BugZilla (e.g., \cite{Antoniol2008, Kochhar2014}), but there is no representative study that considers multiple ITSs in order to understand their influence in the results.

Finally, in \textbf{RQ5}, we analyze how well a model trained with data from various projects can classify samples that belong to a different software system (i.e., one that was not used in training). This is also an open issue in the literature, known as \textit{cross-project classification} \cite{Terdchanakul2017,Zimmermann2010,Wang2016a} and our results, complemented with the ones from the previous RQs, will help developers and researchers in creating models that are able to classify issue reports from projects that are just starting or simply do not contain sufficient information for training purposes.

\begin{figure}[h!]
\centerline{\includegraphics[width=43mm,height=43mm]{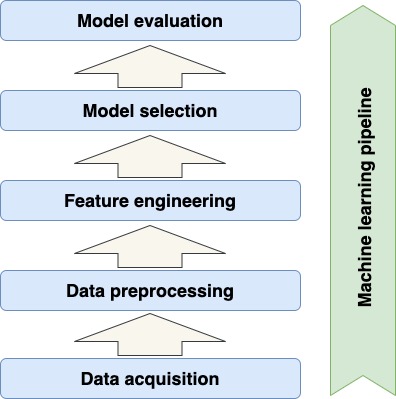}}
\caption{Steps of a typical ML pipeline}
\label{ml_pipeline}
\end{figure}

In the following sections, we discuss the experimental approach followed, which is based on a classical Machine Learning pipeline \cite{hapke2020,Sammut2011}, as illustrated in Figure \ref{ml_pipeline}: \textit{i)} data acquisition, \textit{ii)} data preprocessing, \textit{iii)} feature engineering, \textit{iv)} model selection, and \textit{v)} model evaluation.

\subsection{Data Acquisition}
\label{subsec_data_acquisition}

To answer the research questions, and as a preparatory step, in previous work, we presented the BugHub dataset \cite{Andrade2024}. In this section, we present a brief summary of how it was built (to allow the reader to better understand the whole context and the approach) and define which portion of BugHub is used in the context of this paper. Please refer to the work in \cite{Andrade2024} for further details regarding the dataset. We highlight that the BugHub dataset is being continuously updated, so the sample used this study is just a subset of what will probably exist at the time of reading of the paper.

The goal behind BugHub is to provide an heterogeneous dataset, including a number of software projects of different scales (not only in terms of size, but also in terms of reported issues), different natures (i.e., that represent different types of applications), built using diverse and popular programming languages, and supported by distinct ITSs. Obviously, regardless of the software projects selected, bug reports should be publicly available (which is not the case for many projects). 

By using Google Search and mostly going through known open-source communities (e.g. Apache Software Foundation and Mozilla Foundation), we were able to collect 661,431 
issue reports, a number much higher than what is generally found in related work (see Table I). During the process, we identified several projects that had quite a short amount of issue reports, and we ended up excluding them from this study, with our threshold for inclusion being set at a minimum of 2,000 issue reports 
). We aimed at a larger number than what is usually used in related work (please refer to Table 1 for details) due to the fact that we had the intention of performing statistical tests, which is not frequently the case in related work. The collected reports fit into the following two types: \textit{i)} reports that represent \textit{resolved} and \textit{fixed} bugs (i.e., bugs that were acknowledged by developers and for which a fix was made available and is already integrated into the project);  and \textit{ii)} reports that have been revised by developers and actually represent non-bugs, i.e., they refer to other issues such as new feature or improvement requests, documentation changes, or simply correspond to mistakes made by the person who reported the issue. A detailed list of the projects with the respective data collection date is at \cite{SupMaterials2022}
. 

Overall, the issue reports used in this paper refer to 52 distinct software projects, including relatively small ones, with a few thousand lines of code, to larger ones on the scale of millions of lines of code. The set of projects is based on 10 different programming languages, with 32 of them being managed in GitHub, 15 in Jira and the remaining 5 in BugZilla. Although there are a few other options regarding the ITSs (e.g., \texttt{bugs.launchpad.net}), the ones selected are very popular nowadays, are known to have been used in previous works (e.g., \cite{Kallis2019, Pandey2017, Sohrawardi2014}), and also offer the convenience of an API to access the reports.

Table \ref{tab_projects} presents an overview of the dataset used in this paper, considering the main characteristics of the projects and also identifying which projects are used to answer each RQ. As we can see, for RQ1 and RQ2 all projects are used. For RQ3 the main idea is to select projects written in popular programming languages, namely Java, Python, PHP, JavaScript, and C/C++ and understand the influence of the programming language in classification effectiveness. In practice, we randomly selected 5 projects for each of those 5 languages. Similarly, for RQ4, where we aim to study the influence of the issue tracking system on classification effectiveness, we randomly selected projects from the three different systems, all written in the same programming language (to discard possible influence from the language). Finally, for RQ5 (cross-project classification), we consider five projects written in the same language and using the same issue tracking system (ITS) to eliminate these factors from potentially influencing the results. Further details regarding the decisions made for project selection can be found in Section 4. Table \ref{tab_projects} also includes a column named \textit{OPTM}, which indicates the projects used for hyperparameter optimization (see Section \ref{subsec_feat_engineering} for the corresponding details).





\begin{adjustwidth}{.5cm}{.5cm}\centering\begin{threeparttable}[t]
\caption{Projects characterization.}
\label{tab_projects}
\scriptsize
\begin{tabular}{lrrrrrrrrrrrr}
\toprule
\rotf{\textbf{ITS}} & 
\rotf{\textbf{PROJECT}} & 
\rotf{\textbf{MAIN PL}} & 
\rotf{\textbf{RQ1}} & 
\rotf{\textbf{RQ2}} & 
\rotf{\textbf{RQ3}} & 
\rotf{\textbf{RQ4}} & 
\rotf{\textbf{RQ5}} & 
\rotf{\textbf{OPTM}} & 
\rotf{\textbf{ISSUES}} & 
\rotf{\textbf{BUGS}} & 
\rotf{\textbf{NON-BUGS}} 
\\ \midrule
GitHub &Elasticsearch &Java & \cmark & \cmark & \cmark & & \cmark & \cmark &20,026 &9,605 &10,421 \\
GitHub &Netty &Java & \cmark & \cmark & \cmark & & \cmark & &3,757 &2,638 &1,119 \\
GitHub &Bazel &Java & \cmark & \cmark & \cmark & & \cmark & &3,283 &2,110 &1,173 \\
GitHub &Spring Boot &Java & \cmark & \cmark & \cmark & & \cmark & \cmark &6,766 &2,892 &3,874 \\
GitHub &Spring Framework &Java & \cmark & \cmark & \cmark & & \cmark & \cmark &12,734 &4,440 &8,294 \\
GitHub &Terraform &C/C++ & \cmark & \cmark & & & & \cmark &11,458 &5,600 &5,858 \\
GitHub &Godot &C/C++ & \cmark & \cmark & \cmark & \cmark & & &23,727 &21,105 &2,622 \\
GitHub &Tensorflow &C/C++ & \cmark & \cmark & \cmark & \cmark & & &6,546 &4,912 &1,634 \\
GitHub &QGis &C/C++ & \cmark & \cmark & \cmark & \cmark & & &24,080 &20,543 &3,537 \\
GitHub &Electron &C/C++ & \cmark & \cmark & \cmark & \cmark & & &4,454 &3,330 &1,124 \\
GitHub &Radare2 &C/C++ & \cmark & \cmark & \cmark & \cmark & & \cmark &2,749 &1,213 &1,536 \\
GitHub &Nixpkgs &Nix & \cmark & \cmark & & & & &7,405 &5,819 &1,586 \\
GitHub &VSCode &TypeScript & \cmark & \cmark & & & & &31,651 &19,096 &12,555 \\
GitHub &Kibana &TypeScript & \cmark & \cmark & & & & &13,680 &11,461 &2,219 \\
GitHub &Roslyn &C\# & \cmark & \cmark & & & & &10,248 &8,290 &1,958 \\
GitHub &Pandas &Python & \cmark & \cmark & \cmark & & & &9,643 &7,193 &2,450 \\
GitHub &NumPy &Python & \cmark & \cmark & \cmark & & & \cmark &6,126 &4,656 &1,470 \\
GitHub &Salt &Python & \cmark & \cmark & \cmark & & & &12,440 &9,597 &2,843 \\
GitHub &SciPy &Python & \cmark & \cmark & \cmark & & & \cmark &4,620 &3,097 &1,523 \\
GitHub &Weblate &Python & \cmark & \cmark & \cmark & & & &2,583 &1,200 &1,383 \\
GitHub &Bootstrap &JavaScript & \cmark & \cmark & \cmark & & & \cmark &3,260 &1,153 &2,107 \\
GitHub &AngularJS &JavaScript & \cmark & \cmark & \cmark & & & &3,570 &2,358 &1,212 \\
GitHub &Chart.js &JavaScript & \cmark & \cmark & \cmark & & & &3,601 &2,357 &1,244 \\
GitHub &Strapi &JavaScript & \cmark & \cmark & \cmark & & & &2,581 &1,456 &1,125 \\
GitHub &Material-ui &JavaScript & \cmark & \cmark & \cmark & & & &5,395 &3,562 &1,833 \\
GitHub &Serverless &JavaScript & \cmark & \cmark & & & & &2,643 &1,611 &1,032 \\
GitHub &OwnCloud Core &PHP & \cmark & \cmark & \cmark & & & &12,093 &9,076 &3,017 \\
GitHub &CakePHP &PHP & \cmark & \cmark & \cmark & & & \cmark &5,694 &2,584 &3,110 \\
GitHub &Yii2 &PHP & \cmark & \cmark & \cmark & & & &4,439 &2,489 &1,950 \\
GitHub &Symfony &PHP & \cmark & \cmark & \cmark & & & &16,759 &11,602 &5,157 \\
GitHub &NextCloud Server &PHP & \cmark & \cmark & \cmark & & & &15,392 &10,821 &4,571 \\
GitHub &Kubernetes &Go & \cmark & \cmark & & & & &27,128 &19,934 &7,194 \\
Jira &Mesos &C/C++ & \cmark & \cmark & & \cmark & & &4,978 &2,757 &2,221 \\
Jira &MariaDB Server &C/C++ & \cmark & \cmark & & \cmark & & &11,746 &9,855 &1,891 \\
Jira &Impala &C/C++ & \cmark & \cmark & & & & &5,230 &3,793 &1,437 \\
Jira &Traffic Server &C/C++ & \cmark & \cmark & & \cmark & & &3,114 &2,042 &1,072 \\
Jira &MongoDB Server &C/C++ & \cmark & \cmark & & \cmark & & &28,641 &13,730 &14,911 \\
Jira &Thrift &C/C++ & \cmark & \cmark & & \cmark & & &3,389 &2,195 &1,194 \\
Jira &Cassandra &Java & \cmark & \cmark & & & & &8,810 &5,580 &3,230 \\
Jira &Hadoop Common &Java & \cmark & \cmark & & & & &7,491 &4,894 &2,597 \\
Jira &Hadoop HDFS &Java & \cmark & \cmark & & & & &5,785 &3,712 &2,073 \\
Jira &Ignite &Java & \cmark & \cmark & & & & &6,908 &3,735 &3,173 \\
Jira &Kafka &Java & \cmark & \cmark & & & & &4,918 &3,336 &1,582 \\
Jira &Lucene Core &Java & \cmark & \cmark & & & & \cmark &5,880 &2,705 &3,175 \\
Jira &Qpid &Java & \cmark & \cmark & & & & &5,931 &3,890 &2,041 \\
Jira &Solr &Java & \cmark & \cmark & & & & &6,437 &3,619 &2,818 \\
Jira &Spark &Scala & \cmark & \cmark & & & & &1,4554 &8,142 &6,412 \\
BugZilla &Mozilla Core &C/C++ & \cmark & \cmark & & \cmark & & &164,708 &128,608 &36,100 \\
BugZilla &Sea Monkey &C/C++ & \cmark & \cmark & & \cmark & & &9,946 &8,765 &1,181 \\
BugZilla &Firefox &C/C++ & \cmark & \cmark & & \cmark & & &25,423 &18,607 &6,816 \\
BugZilla &Mozilla NSS &C/C++ & \cmark & \cmark & & \cmark & & &6,493 &4,144 &2,349 \\
BugZilla &Thunderbird &C/C++ & \cmark & \cmark & & \cmark & & &10,518 &7,931 &2,587 \\
\multicolumn{9}{r}{\textbf{AVERAGE}} &\textbf{12,720} &\textbf{8,843} &\textbf{3,877} \\
\multicolumn{9}{r}{\textbf{SUM}} &\textbf{661,431} &\textbf{459,840} &\textbf{201,591} 
\\
\bottomrule
\end{tabular}
\end{threeparttable}\end{adjustwidth}


\subsection{Data Preprocessing}
\label{subsec_data_preprocessing}

We applied several preprocessing techniques to the collected issue reports, as a way to assure the data quality. Namely, the preprocessing techniques frequently used in Natural Language Processing problems, starting with the conversion of plain text to lowercase, the removal of punctuation and isolated characters, as these do not bring value to training and can even impair it \cite{Babanejad2020,Uysal2014}. Next, we removed general stop-words (e.g., \textit{a}, \textit{the}, etc), with exception of a few meaningful terms, such as "not", which previous work has suggested being important in this context \cite{Antoniol2008}. Finally, we used lemmatization \cite{thanaki2017,vajjala2020} to convert words into their root form. We opted for lemmatization, instead of stemming, considering that the latter reduces words to their common root by removing or replacing word suffixes (e.g., “loading” is stemmed as “load”), while the former identifies the inflected forms of a word and returns its base form (e.g. “better” is lemmatized as “good”) \cite{Korenius2004}. All the details regarding the various configurations used are available at \cite{SupMaterials2022}.

\subsection{Feature Engineering}
\label{subsec_feat_engineering}

All the reports used in this work have been tagged as `resolved' and `fixed', which basically means that developers confirmed the presence (or absence) of an issue. Regarding the automatic classification itself, some studies use attributes extracted from both structured and unstructured fields (e.g., \cite{Kochhar2014, Kukkar2018}). In this work, we are solely interested in unstructured data, particularly in the text associated with the issue report, i.e., report title and full textual description, as these data are generally available across different ITSs or projects. The same does not happen with some structured data, such as characteristics of the individual reporting the bug, or may be available with clear differences across projects, like distinct values for characterizing the severity of a bug. Such cases impair the definition of a general model for classifying bug reports.

We followed a Bag of Words approach with features being generated by \textit{term frequency–inverse document frequency} (TF-IDF) \cite{Manning2008}. Although some authors use only term frequency (e.g., \cite{Chawla2015}), we consider that adding inverse term frequency can make the feature generation more efficient, as only words that are meaningful for each class are kept, instead of keeping terms that appear more frequently in the whole corpus of training data. This is corroborated by previous works, which show that TF-IDF is a suitable choice for this particular context \cite{Kochhar2014,Sohrawardi2014,Limsettho2014,Kukkar2018}.

A common problem related to Bag of Words approaches is the curse of dimensionality \cite{Aggarwal2015,Verleysen2005}, a phenomenon that inevitably arises as the total dimension is the vocabulary size. In other words, it often occurs because the number of terms found in the training corpus is too big. To avoid this issue, an adequate feature selection process must be employed. Common options are forward  selection and backward elimination \cite{Kotu2019}. In this work, we actually opted for using the chi-squared method \cite{Novakovic2011}, due to the fact that it was successfully used by other authors in a context similar to our own and considering that it requires less processing power in comparision to the remaining ones 
\cite{Kukkar2018,Terdchanakul2017}. In practice, the chi-squared method will keep only the \textit{n} features which individually lead to better classification performances, with \textit{n} being a parameter chosen by the user. We analyze the best suitable value for \textit{n} in the scope of our first research question, as detailed in Section \ref{section_res_rq1}. 

\subsection{Models Training}
\label{subsec_model_selection}

To create the models, we consider different types of classical ML algorithms for supervised learning that were previously used in the context of issue report classification \cite{Lopes2020,Kochhar2014,Sohrawardi2014,Terdchanakul2017}. We selected the k-Nearest Neighbor classifier (k-NN) \cite{Cunningham2021}, known to be a lazy learner; Naïve Bayes (NB) \cite{Kotu2019}, which is a Bayesian classifier; Support Vector Machine (SVM) \cite{Cortes1992}, a maximal-margin classifier; Random Forest (RF) \cite{Breiman2001}, an ensemble classifier; and Logistic Regression (LR) \cite{Aggarwal2015}, a probabilistic approach. 

We start our study in RQ1 with the \textit{Logistic Regression} algorithm, as it does not require any particular tuning of hyperparameters (which would lead to a massive number of multiple combinations to be explored, which is out of the scope of this work). Based on the results of RQ1, namely which data to use (i.e., report titles or descriptions) and the number of features to adopt, we carry out an analysis of the classification effectiveness for all the selected ML algorithms in RQ2, producing evidence suggesting that, in our scenario, SVM, LR and RF are suitable algorithms to handle classification of issue reports. Based on such evidence, the discussions on RQ3, RQ4 and RQ5 are based on results obtained with SVM. Nevertheless, the complete set of results for all algorithms (i.e., k-NN, LR, NB, and RF) are available at \cite{SupMaterials2022}.

Configuring the models mainly involves selecting the best hyperparameter values. For this purpose, we performed a grid search \cite{Syarif2016} over the 10 projects where we observed the best classification performance in RQ1. We opted to use a subset of the projects to avoid the large cost of running a grid search for the complete dataset. In practice, a grid search was run 10 times for each of the 10 projects and for each of the five classifiers, and we ended up selecting the set of parameter values that appeared more frequently. 
Table \ref{tab_params_classifiers} presents the tested and chosen values for the hyperparameters.









\begin{table}[!htp]\centering
\caption{Parameters used for optimization.}
\label{tab_params_classifiers}
\scriptsize
\begin{tabular}{lrrrr}\toprule
\textbf{Classifier} &\textbf{Param} &\textbf{Tested Values} &\textbf{Selected} \\\midrule
\multirow{3}{*}{LR} &penalty &l2 &l2 \\
&C &0.5, 1.0, 1.5 &1.5 \\
&solver &newton-cg, lbfs, liblinear, sag, saga &newton-cg \\
\multirow{3}{*}{SVM} &kernel &rbf, linear &linear \\
&gamma &1e-3, 1e-4 &1e-3 \\
&C &1, 10, 100, 1000 &100 \\
\multirow{3}{*}{RF} &n\_estimators &25, 50, 100, 200, 300 &200 \\
&criterion &entropy, gini &entropy \\
&min\_samples\_leaf &5, 10, 25, 50 &5 \\
\multirow{5}{*}{KNN} &algorithm &auto, brute &auto \\
&weights &uniform, distance &uniform \\
&leaf\_size &10, 20, 30, 40, 50 &10 \\
&n\_neighbors &3, 5, 7, 9 &9 \\
&p &1,2 &2 \\
\bottomrule
\end{tabular}
\end{table}

\subsection{Models Evaluation}
\label{subsec_model_evaluation}

For the evaluation of the models, we followed a set of general practices:

\begin{enumerate}
    \item \textbf{We used 70\% of the data for training and the remaining 30\% for testing the models}. The samples were randomly picked from the dataset and ensured that the samples used during training were never present in the testing phase \cite{mueller2021,theobald2024}. 
    
    \item \textbf{For training, we used balanced data}, i.e., different classes are present in the same proportion in the data that is used to train the models. We use balanced data during training to avoid creating models that would otherwise tend to classify a given report as being of the majority class, thus negatively affecting the performance of the algorithm \cite{marwala2018,fernández2018}. For this purpose, we used undersampling \cite{Yap2014}. Although other techniques exist, undersampling offers better run times and, by randomly executing each experiment several times, we minimized the risk of losing important information in the majority class.
    
    \item \textbf{For testing, we used imbalanced data}, i.e., the different classes appear in the original proportion found in the field, as extracted from the source ITSs (exact proportions are shown in Table \ref{tab_projects}). This allows an understanding of the effectiveness of the models in conditions that are representative of real scenarios \cite{he2013,abhishek2023}.
    
    \item \textbf{We executed 30 training/testing repetitions}, using randomly selected data in each one, to study how results may vary. By the end of each repetition, results were stored to allow calculating specific metrics, as discussed in the following sections.
    
    \item \textbf{We analyze all results resorting to visual observation and statistical testing} (e.g., to compare results obtained by different alternatives, such as titles \textit{versus} descriptions). We report average values as in \cite{Fan2020, Terdchanakul2017}, allowing for an analysis in perspective with related studies. To check the requirements for applying parametric tests, we considered the following methods: \textit{i}) a Shapiro–Wilk test \cite{Shapiro1965} to check whether metric values (e.g., F-measure values) follow a normal distribution or not; and \textit{ii}) both Levene’s \cite{Levene1960} and Fligner-Killeen test \cite{Fligner1976} to verify if the variance is similar among the samples being tested. In case the data we are analyzing do not fulfill the assumptions to carry out parametric tests (which, as we will see in Section V, turns out to be all cases), we used a non-parametric method, namely a pairwise version of the \textit{Wilcoxon rank sum test} (also known as \textit{Mann Whitney U test}) \cite{Wilcoxon1963} with \textit{p-values} adjusted by the \textit{Bonferroni} method \cite{hsu1996}. We consider a typical value of \(\alpha=0.05\).
\end{enumerate}

Common metrics for model evaluation are \textit{accuracy}, \textit{precision}, \textit{recall},  and \textit{F-measure}. \textit{Accuracy} refers to the number of correctly predicted examples out of all the examples and is calculated as shown in Equation (\ref{eq_accuracy}). 

\begin{equation}
    \label{eq_accuracy}
    Accuracy = \frac{(TP + TN)}{(TP + FP + TN + FN)}
\end{equation}

In Equation (\ref{eq_accuracy}), \textit{TP} means True Positives, which is the number of instances correctly classified in the positive class; \textit{FP} represents False Positives, which are instances from the negative class misassigned into the positive class; \textit{TN} denotes True Negatives, i.e., examples from the negative class which are correctly classified; and \textit{FN} is the number of False Negatives, i.e., examples from positive class that are incorrectly classified to negative class. Accuracy may be misleading in some situations, as for example when data is highly imbalanced. In such classification problems, algorithms may classify poorly the less frequent class, but still present high accuracy values if the most frequent class is appropriately classified. Other types of metrics should be used as a clearer way of representing classification effectiveness, namely \textit{F-measure} \cite{Japkowicz2011}. It is calculated as the harmonic mean of \textit{precision} and \textit{recall}. The formulas to calculate precision and recall are defined in (\ref{eq_precision}) and (\ref{eq_recall}), respectively. F-measure is should be computed as shown in (\ref{eq_f-measure}).

\begin{equation}
    \label{eq_precision}
    Precision = \frac{TP}{TP + FP}
\end{equation}

\begin{equation}
    \label{eq_recall}
    Recall = \frac{TP}{TP + FN}
\end{equation}

\begin{equation}
    \label{eq_f-measure}
        F_1 = 2 * \frac{Precision * Recall}{Precision + Recall}
\end{equation}

As we are dealing with scenarios where the classes are not balanced in the testing sets, \textit{we resort to F-measure as a metric to evaluate the models}, 
following the general ML pipeline described before. When applicable, additonal  steps taken for each specific RQ are described in following sections. The detailed results, dataset, and all code necessary for replicating the experiments are available at \cite{SupMaterials2022}. 

\section{Dimensionality and Issue Report Content (RQ1)} 
\label{section_res_rq1}

This section presents the experimental results for our first RQ. As mentioned, the first objective is to understand which content (i.e., titles or descriptions) is more useful for classifying issue reports. For this, we resorted to all data available in the 52 projects gathered from GitHub, Jira and Bugzilla ITSs (see in Table \ref{tab_projects}). For each project, we used the Logistic Regression classifier to execute the train/test repetitions (see Section \ref{subsec_model_evaluation}), as this algorithm does not require any particular tuning of hyperparameter values.

 
We selected the following values to be used for the number of dimensions: \{50, 100, 150, 200, 250, 300, 350, 400, 450, 500\}. We selected these values, as some of the projects we are using do not contain much more than 1,000 examples per class (i.e., \textit{bug} and \textit{non-bug}), thus using more than 500 dimensions is not reasonable (e.g., due to high overfitting chances). We then ran the 30 train/test repetitions, first using report titles and then the descriptions. For each of these two types of information, we obtained a total of 15,600 outcomes, considering the 10 different dimension values, 52 projects and 30 train/test repetitions per project. 

\begin{figure}[!h]
\centerline{\includegraphics[width=75mm]{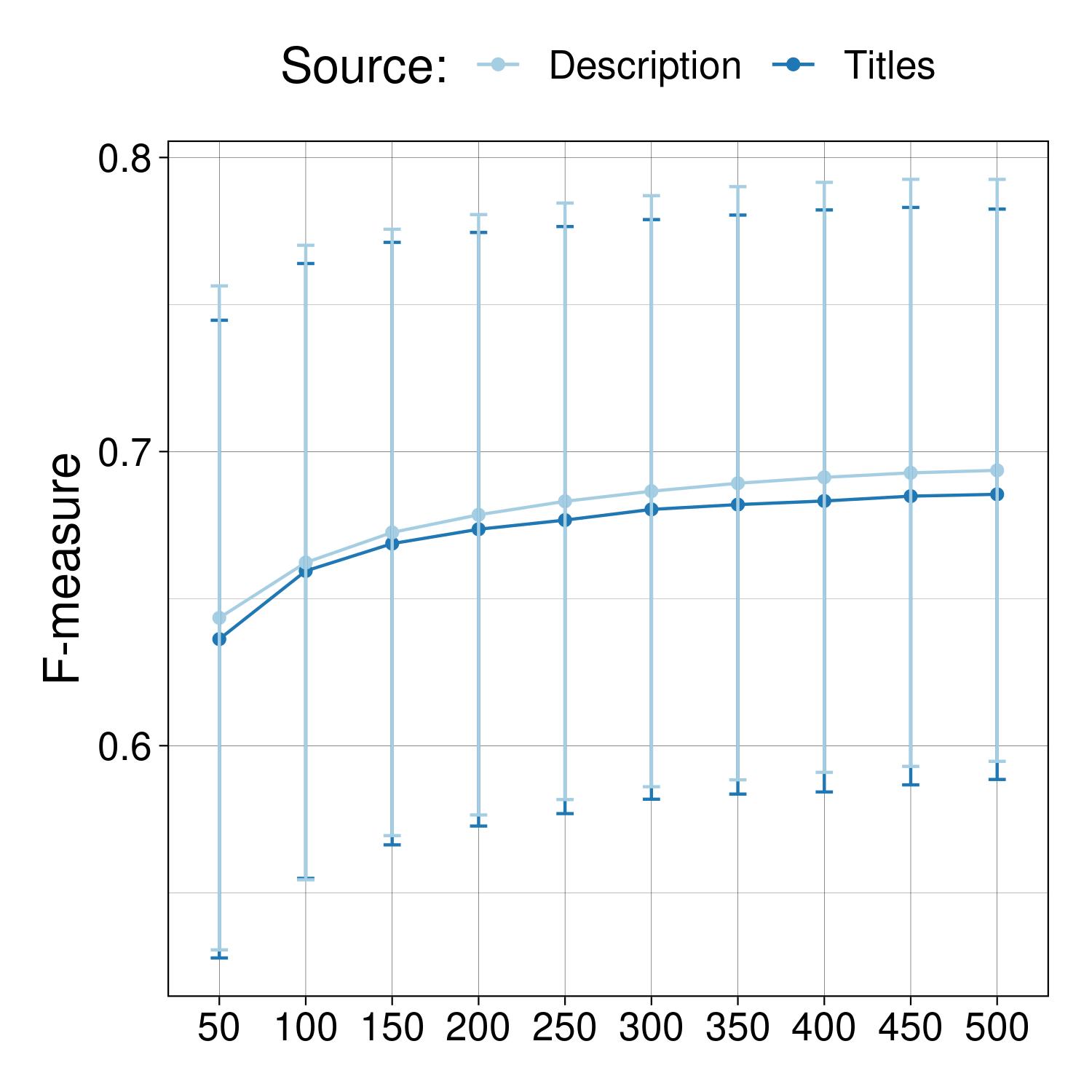}}
\caption{Mean F-measure by number of dimensions.}
\label{lp_f1_by_dim_summ_vs_desc}
\end{figure}

Figure \ref{lp_f1_by_dim_summ_vs_desc} shows the results obtained for the different dimension values, where we can see that the average F-measure gradually increases with the number of dimensions, with more expressive value changes in the 50 to approximately 250 dimensions range. When the dimension size exceeds 250, the differences in the F-measure seem smaller for both titles and descriptions. Figure \ref{lp_f1_by_dim_summ_vs_desc} also shows some difference between the F-measure when using titles and when using descriptions, which is more expressive in the cases of 400, 450 and 500 dimensions (although the maximum difference is never larger than 1\%). Also noticeable are the error bars, which are large since the F-measure values largely vary from project to project (e.g., considering 250 dimensions, the mean F-measure of the Firefox project is around 0.57, while Spring Boot reaches about 0.83). 


After visually analyzing the data, we statistically tested whether the F-measure values vary significantly with the increasing number of dimensions. We verified if the data allowed us to carry out parametric tests, which was not the case as we are dealing with several cases were the data do not follow a normal distribution. Also, the variance is not similar among the classes we were testing. We then proceeded with  non-parametric tests, as discussed in Section \ref{subsec_model_evaluation}. We found out that, \textbf{for both titles and descriptions, there are no statistically significant changes between the performance of models generated using 250 dimensions and those generated with higher dimension values}, as the \textit{p-values} obtained are greater than the predefined significance level \(\alpha=0.05\) 

To understand the reasons behind our findings, we further analyzed the raw data, configuration and outcome of the methods used. When studying the importance of each term in the BoW in what regards to the whole corpus of issue reports, we found out that a small number of words is often associated with high relevance in the feature selection process carried out to create the models. By analyzing the TF-IDF weights associated with individual words, we noticed that the average score given to the first 250 terms of the bag-of-words is about one-third higher than the mean value of the weights assigned to words lying in the range that goes from 250 to 500 dimensions. In a random example extracted from our experiments carried out with the AngularJS project, when ranking a set of terms from titles by their TF-IDF scores, we observed that the bi-grams \textit{properly handle} and \textit{implement detach} appear in two of the first positions, with weights of 7.8 and 3.3 respectively. In contrast, the word \textit{group} occurs in position 255, associated with a score of 0.2. This indicates that features used by 250-dimension models have the most valuable terms and adding more attributes would not help in significantly improving performance. Instead, we observed that models from many projects actually suffer performance losses when created with a higher number of attributes, as they become strictly adapted to the training data and lose their ability to perform as expected when dealing with the testing sets (i.e., due to overfitting reasons). The outcome of this first part of RQ1 may benefit future works in the area, as finding the right number of dimensions to use is time- and resource-consuming and no empirical evidence is available on such an issue in the literature. 
In addition, it contributes to avoiding the \textit{curse of dimensionality} problem, which is common in BoW approaches.

\begin{figure}[h!]
\centerline{\includegraphics[width=60mm]{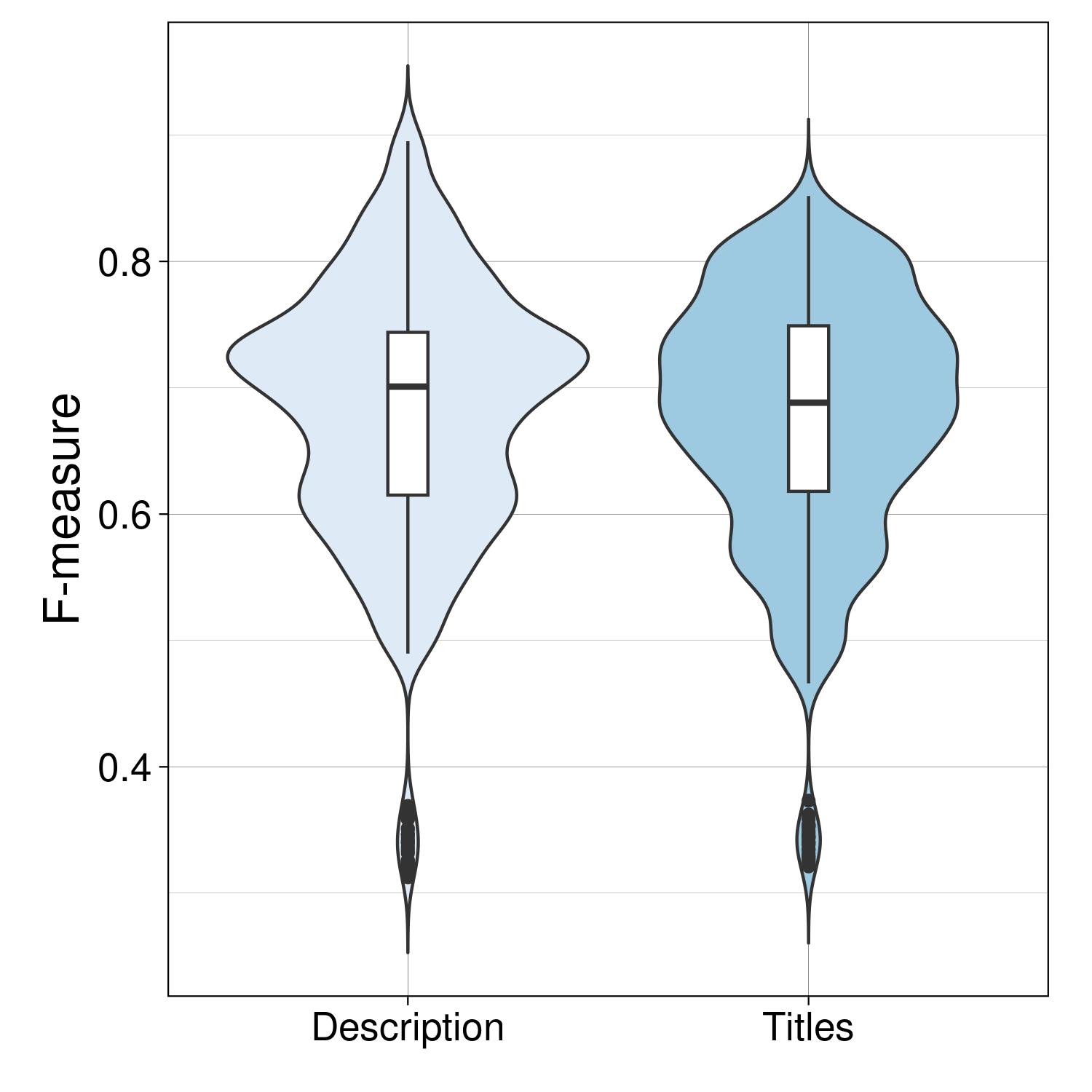}}
\caption{Performance comparison - titles \textit{vs.} descriptions.}
\label{vl_comp_summ_vs_desc}
\end{figure}

Considering 250 dimensions, we then assessed whether there are significant differences between the use of titles and the use of descriptions to train the models. Figure \ref{vl_comp_summ_vs_desc} shows the F-measure values obtained for the Logistic Regression in both cases. In the chart, the shapes are kernel density estimations to show the data distribution. As we can see, the shapes formed by the distribution of the F-measure values are different, although mean values and interquartile seem relatively close. It also suggests that the F-measure values of the experiments using titles may be closer to a normal distribution than those obtained with the descriptions.

To confirm our visual observations, we statistically tested if the F-measure obtained when using titles is significantly different from the one obtained when using the descriptions. For this purpose, our hypothesis testing revealed a \textit{p-value} of 0.1851, which is greater than the defined \(\alpha\) level, showing no evidence to confirm that the effect of using either titles or descriptions indeed makes a difference. Thus, our conclusion is as follows: 
 
 \begin{tcolorbox}[colback=gray!2!white,colframe=gray!75!gray!50!]
    \textbf{Response to RQ1:} There are no statistically significant differences in the F-measure values obtained with Logistic Regression models trained with either titles or descriptions of issue reports.
\end{tcolorbox}

The explanation for such a finding resides in the fact that the low number of terms present in the title is connected to high importance in the decision process, while most words in the description add poor or even no value to the models, in terms of classification robustness. While a full description may be useful for developers (e.g., for helping them to correct the issues), for the purpose of identifying if an issue represents a bug or not, it might not be the most reasonable choice, considering that more processing power is required to handle extensive texts like those usually present in a typical description of an issue report. By looking specifically at the weights given by TF-IDF to each term, it is noticeable that, on average, the scores for terms in the titles are about one-fifth higher than those of descriptions. In other words, the strong weighted words in titles tend to compensate for the performance losses caused by their smaller number and diversity in comparison with the descriptions, eliminating statistically significant differences in the performance of models. 

To further understand the performance differences between models built from descriptions \textit{vs.} models created by using titles, we also analyzed results considering the issue tracking system associated with the projects. We noticed that models from projects hosted by both Jira and BugZilla, reach higher performance if built using titles, rather than descriptions. On the opposite side, models created with projects from GitHub tend to have higher performance when using descriptions. This may occur because the overall description length of GitHub issue reports is on average 41\% higher than those of BugZilla and about 14\% greater than those of Jira. Due to their completeness, descriptions are to some extent capable of producing better results when using projects from GitHub, although our hypothesis testing showed that the performance improvement is not significant in comparison with models created by using titles. As titles are shorter than descriptions, using them is a reasonable choice to reduce the amount of processing power required to create models to carry out automatic classification of issue reports. 


Supported by the evidence found during the RQ1 experiments, we conducted the remaining experiments in this work based on the use of \textit{report titles} and \textit{250 dimensions}, allowing us to save computational resources while not sacrificing the effectiveness of the models.

\section{Performance of the Classifiers (RQ2)} 
\label{section_res_rq2}

To understand if there are significant differences in performance achieved by different classifiers, we resorted to the full dataset and to the  five algorithms previously introduced in Section \ref{sec_study}: NB, RF, LR, SVM and KNN. 
Considering that we have 52 projects, five algorithms, and we carry out 30 train/test repetitions for each project, per algorithm, we end up with 7,800 resulting F-measure values.

\begin{figure}[b!]
\centerline{\includegraphics[width=65mm]{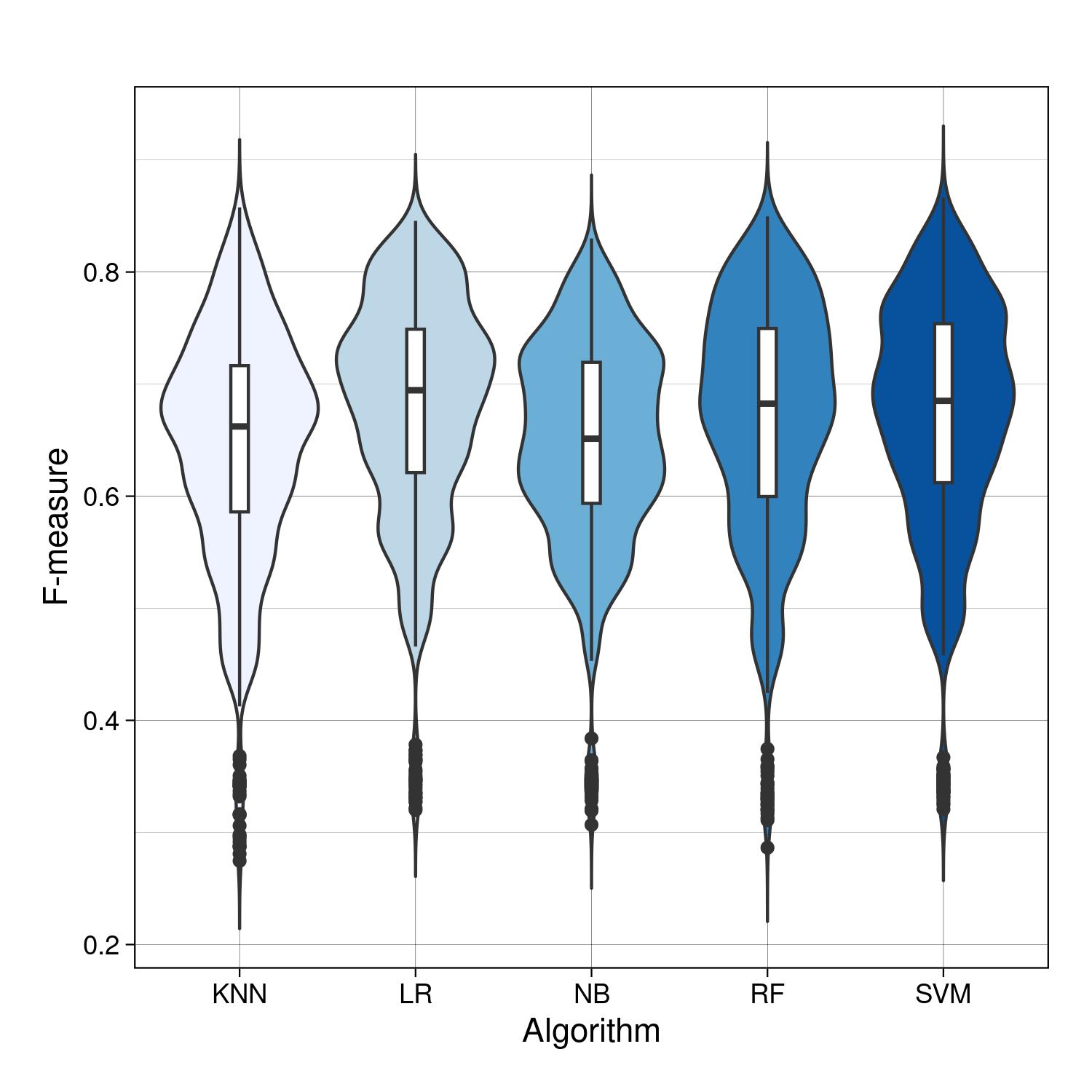}}
\caption{F-score comparison between NB, LR, RF, SVM and KNN algorithms.}
\label{vl_algorithms}
\end{figure}

A visual comparison between the performance of the five classifiers is presented in Figure \ref{vl_algorithms}. As we can see, the shapes resulting from the F-measure distributions seem to be quite similar across all algorithms. Despite such similarity, the chart also shows that LR, RF and SVM present slightly higher performance when compared with NB and KNN classifiers. A more detailed comparison is provided in Table \ref{tab_comp_classifiers}, where we can see the 30 runs average F-measure values reached by each classifier per programming language. 
As it can be noticed by looking at the table, there are cases in which the differences between the classifiers' performance seem relatively small, as for example in C/C++ projects. In contrast, models based on issue reports from projects in PHP tend to indicate a more expressive difference between two groups of classifiers: \textit{i)} LR, RF and SVM; and \textit{ii)} RF and KNN.

In terms of F-score, RF and SVM achieved an average F-measure of 0.67 with LR reaching 0.68. NB and K-NN reached an average F-score of 0.65. We tested whether there were statistically significant differences in the F-measure values obtained, by comparing the values obtained for each algorithm with all the remaining four. The results pointed out the existence of statistically significant differences between two groups of algorithms. Namely, the results obtained with RF, SVM, and LR were found to be statistically different from the ones achieved by NB and K-NN, with the corresponding \textit{p-values} being lower than the value of $alpha$ set to $0.05$.
This leads us to the following conclusion:

\begin{tcolorbox}[colback=gray!2!white,colframe=gray!75!gray!50!]
    \textbf{Response to RQ2:} The F-measure values reached by SVM, LR, and RF are not significantly different. NB and KNN show significantly (lower) performance when compared against the former set of algorithms.
\end{tcolorbox}









\begin{table}[!htp]\centering
\caption{F-score comparison between different classifiers per programming language.}
\label{tab_comp_classifiers}
\scriptsize
\begin{tabular}{lrrrrrr}\toprule
\textbf{Language} &\textbf{NB} &\textbf{LR} &\textbf{RF} &\textbf{SVM} &\textbf{KNN} \\\midrule
C/C++ &0.61 &0.62 &0.61 &0.62 &0.59 \\
C\# &0.53 &0.55 &0.53 &0.50 &0.50 \\
Go &0.70 &0.72 &0.72 &0.72 &0.70 \\
Java &0.68 &0.72 &0.72 &0.72 &0.69 \\
JavaScript/TypeScript &0.66 &0.69 &0.68 &0.69 &0.67 \\
Nix &0.60 &0.61 &0.58 &0.61 &0.57 \\
PHP &0.62 &0.68 &0.67 &0.68 &0.64 \\
Python &0.71 &0.74 &0.73 &0.72 &0.69 \\
Scala &0.68 &0.73 &0.73 &0.72 &0.70 \\
\bottomrule
\end{tabular}
\end{table}

Our analysis of the results confirms that SVM, as well as LR and RF outperform the remaining classifiers. In perspective with the related work, to the best of our knowledge and at the time of writing, there is no clear empirical evidence on the performance of SVM, LR, RF, KNN and NB classifiers, compared to each other, despite they appear commonly together in this context. Moreover, previous studies where performance comparisons are made were conducted with limited datasets, commonly including about 5k issue reports, in which only one ITS is normally considered. These datasets may not be representative enough to allow the reaching of solid conclusions. Most of them use a small number of projects, which often range from 3 to 7. Also, most common projects are written in Java, while other programming languages are hardly considered. Thus, our findings in RQ2 provide important guidance for future research on automatic classification of issue reports, particularly regarding model selection, as we pointed out suitable options of algorithms to be used.

Based on the conclusions obtained during the experiments associated with RQ2, we opted to focus our efforts on SVM for the analysis of results from the tests made in the scope of the next research questions, as this classifier is known to be widely used for automatic classification of issue reports\cite{KumarNagwani2012,Kochhar2014,Sohrawardi2014,Pandey2017,Panda2019}.


\section{Programming Languages (RQ3)} 
\label{section_res_rq3}

Our goal in RQ3 is to understand if the main programming language in which the project is written can be a factor that influences the performance of the models when classifying issue reports. Indeed, reports may have content that is coupled to structures or mechanisms of certain programming languages, such as specific exceptions being raised, or memory pointers being misused. Such aspects may influence the quality of the classification. 

We focus on five different and popular programming languages, namely, Java, Python, PHP, JavaScript, and C/C++ (the latter one is a combination of two languages that share many similarities) and selected 25 different projects to create the models (i.e., 5 per programming language, as previously shown in Table \ref{tab_projects}). All projects selected for our experiments in this RQ are supported by a single ITS (i.e., GitHub), as using projects from more than one tracking system could add bias to the results.

For each programming language, we created a dataset of 8,000 randomly selected reports (i.e., 1,600 issue reports from each of the 5 projects written in that particular programming language). We use the same amount of reports per project to avoid biasing the results towards the projects with more issues. At the same time, this amount cannot exceed the number of reports of the smallest project (the resulting set would not possess an equal number of reports per project) and we opted for it to be actually lower than half of that number to allow for sufficient randomization across multiple runs. We then divided the set into training and testing sets, in the usual proportions and balancing rules as presented in Section \ref{subsec_model_evaluation}.

\begin{figure}[!h]
\centerline{\includegraphics[width=65mm]{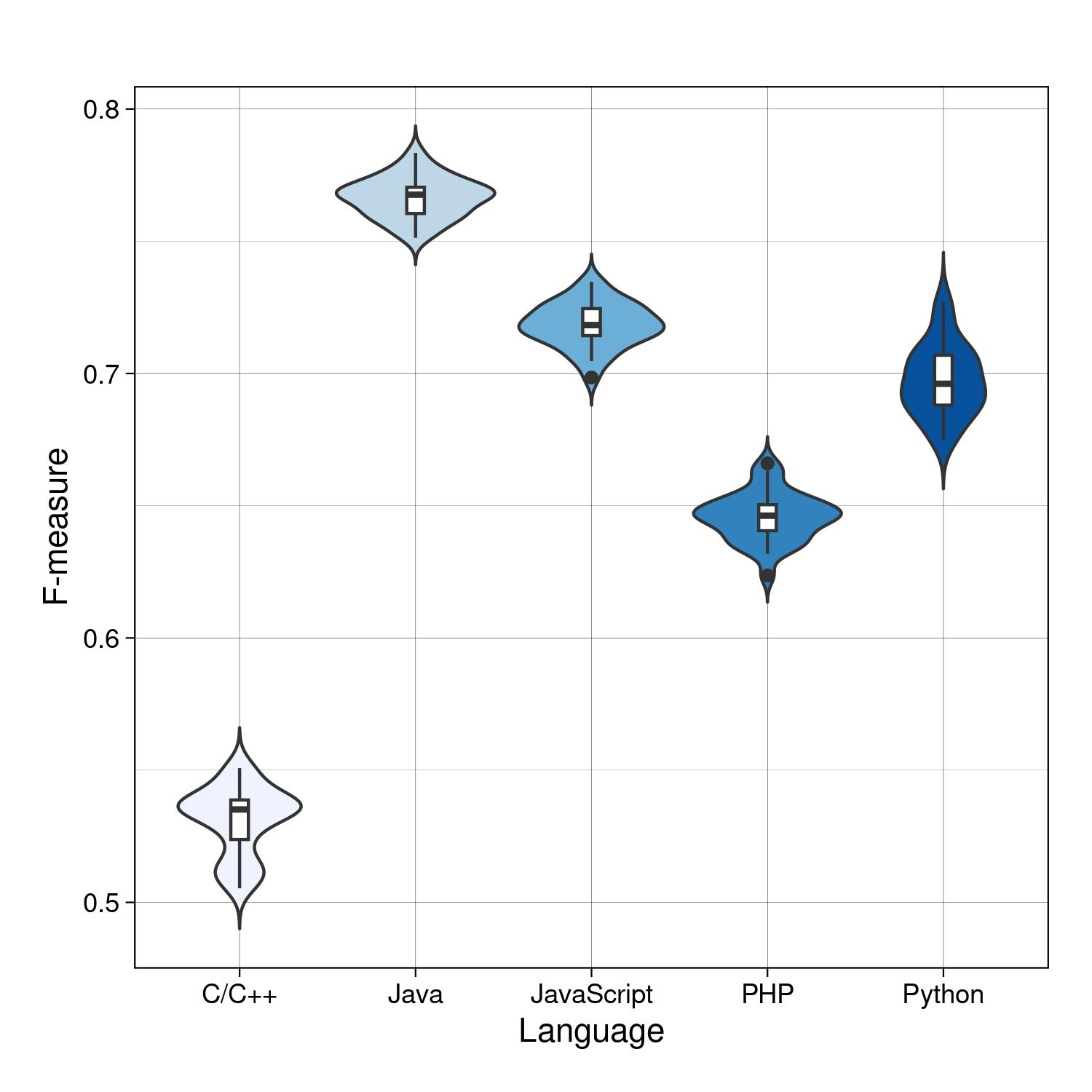}}
\caption{F-score comparison between programming languages.}
\label{vl_languages}
\end{figure}

Figure \ref{vl_languages} shows the distribution of the F-measure values obtained by SVM for each of the programming languages considered in RQ3. As we can see, there are clear visual differences between the F-measure values. Java is generally associated with higher F-measure values, followed by JavaScript and Python, which however seem to present close results. The chart also shows that models perform poorly when handling PHP and also C/C++ projects, which are associated with the worst results. 


We statistically tested if the F-measure is significantly different depending on the programming language. The results of the tests confirmed that the average F-score values that were observed for C/C++ (0.53), Java (0.77), JavaScript (0.72), PHP (0.65) and Python (0.7) are statistically significant differences for all cases (i.e., when comparing the results of each programming language with all remaining) 
, allowing us to reach the following conclusion:

\begin{tcolorbox}[left=1pt,top=1pt,bottom=1pt,right=1pt,colback=gray!2!white,colframe=gray!75!gray!50!]
\textbf{Response to RQ3:}  The classification performance of the SVM algorithm is significantly different when classifying projects that differ on the main programming language.

The F-measure values obtained with SVM for projects in which the main programming language is different are associated with statistically significant differences.

\end{tcolorbox}

It is worthwhile mentioning that the classification differences observed were consistent across not only SVM, but also with the remaining algorithms. 
We further tried to understand the reasons behind the observed performance variations and began by analyzing the terms selected to build the models and their TF-IDF scores, but found no evidence to support such differences. Instead, we discovered that the degree of imbalance in the testing data (i.e., the proportion of bugs \textit{vs.} non-bugs) is strongly associated with the performance of the models. As previously observed, models built based on issues retrieved from Java projects reach higher performance, whereas the ones built based on C/C++ project issues are linked to lower performance values. In our dataset, bug-related issue reports from Java projects represent 47\% of the total amount, while non-bug-related are about 53\% of the examples. In the issue reports referring to C/C++ projects, the proportion of bug-related examples is about 83\% and non-bug-related reports appear in only 17\% of the cases.

Based on the hypothesis of the imbalance of the data being a reason for the performance differences, we executed a set of additional experiments 
For this, we trained a set of models 
using different proportions of bug- and non-bug-related issue reports from Java projects, and analysed the results. Table \ref{tab_lang_complementary} shows how the F-measure varies with random proportions of bug- and non-bug-related issue reports (columns \textit{Mean Prop. Bugs} and \textit{Mean Prop. Non-bugs}). As we can see, the performance tends to increase as the classes become more balanced. Statistical testing showed that when applying the same proportion observed in C/C++, as shown in the first line of Table \ref{tab_lang_complementary}, F-measure values are significantly inferior to the ones resulting from models trained and tested with data from the C/C++ projects themselves.
Yet, we highlight that the degree of imbalance is an intrinsic characteristic of the projects, which tends to be similar when the same programming language is considered. This helps understanding the performance variations observed in RQ3, but especially helps in guiding future studies aiming at creating more generic models that are effective in classifying issue reports, regardless of the programming language involved. 









\begin{table}[!htp]\centering
\caption{F-score variation of Java models with different class proportions in the testing set.}
\label{tab_lang_complementary}
\scriptsize
\begin{tabular}{lrrrr}\toprule
\textbf{Language} &\textbf{Mean Prop. Bugs} &\textbf{Mean Prop. Non-bugs} &\textbf{Mean F-score} \\\midrule
Java &83\% &17\% &0.48 \\
Java &70\% &30\% &0.63 \\
Java &60\% &40\% &0.70 \\
Java &50\% &50\% &0.76 \\
\bottomrule
\end{tabular}
\end{table}

\section{Issue Tracking Systems (RQ4)} 
\label{section_res_rq4}

In RQ4, we aim to analyze whether the ITS being used by the project might influence the performance of models or not. We consider three different ITSs, namely GitHub, Jira and BugZilla. As we found out previously that the programming language is a factor that influences performance, we focused on projects written in a single language. We selected C/C++ as a target, as we have projects coded in this programming language present in all three tracking systems, and considered
a total of 15 projects (i.e., 5 projects per ITS). 

\begin{figure}[h!]
\centerline{\includegraphics[width=65mm]{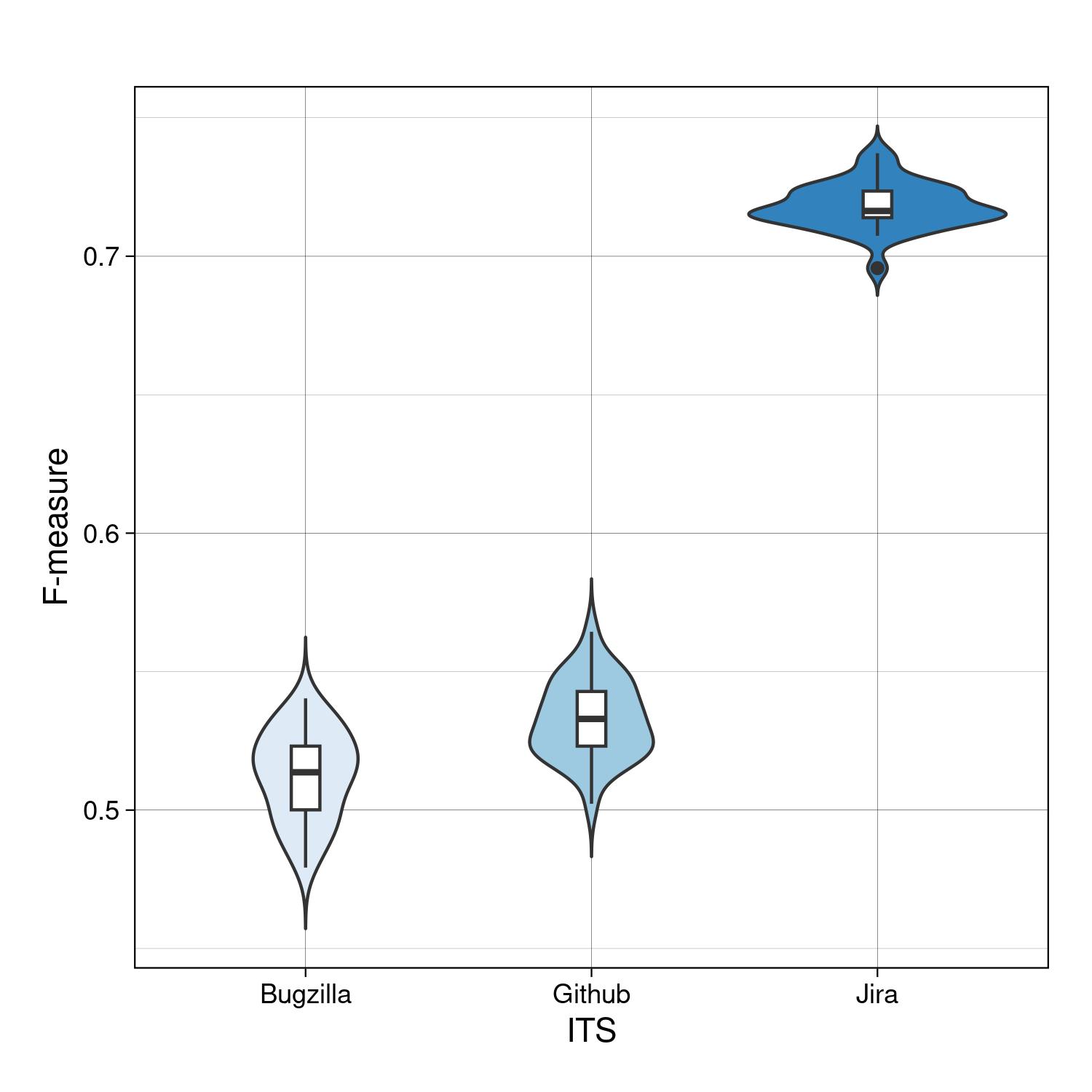}}
\caption{F-score comparison between issue tracking systems.}
\label{vl_tracking_systems}
\end{figure}

Figure \ref{vl_tracking_systems} shows the distribution of the data regarding the F-measure values obtained with SVM and concerning reports from GitHub, Jira, and Bugzilla. As we can see, Jira is clearly associated with the highest F-measure values (0.72), followed by GitHub (0.53) and Bugzilla (0.51), which are both quite distant from the Jira values (i.e., the difference is nearly 20\%). GitHub actually scores higher than Bugzilla but the difference is small (i.e., about 2\%).

After visually observing the results, we proceeded with the statistical tests, to assess if the F-measure values associated with the different ITSs supporting the projects were significantly different. All resulting \textit{p-values} revealed to be lower than the significance level of \(\alpha=0.05\)
, thus supporting the following conclusion:

\begin{tcolorbox}[left=1pt,top=1pt,bottom=1pt,right=1pt,colback=gray!2!white,colframe=gray!75!gray!50!]
    \textbf{Response to RQ4:} 
    The F-measure values obtained with SVM for projects supported by different ITSs are associated with statistically significant differences.
\end{tcolorbox}

It is important to mention that we observed consistent results across all algorithms in the paper(i.e., despite the obvious different absolute values observed with each algorithm). We then proceeded to analyze the factors that could potentially be related to the observed performance variations. We noticed that similarly to what happens in RQ3, final class distributions are different depending on the ITS which hosts the projects. As shown above, Jira is associated with models that achieve the best performance, and BugZilla with the models that reach the lower F-measure values. In the data we used for our experiments, the proportion of bug- and non-bug-related issue reports from Jira projects is 61\% and 39\% respectively, while software systems that use BugZilla present an overall proportion of 78\% bugs and 22\% non-bug reports. To confirm if the performance variation observed in the results from RQ4 is related to the degree of imbalance found in the issue reports from projects hosted by each tracking system, we executed a set of additional experiments by using the exact same procedure adopted in RQ3, but resorting to different class proportions in the testing set. We were able to confirm that the performance of the models is lower when classes are not balanced in the testing set. Furthermore, we noticed that when models from Jira projects are tested with data in which the proportion of bug- and non-bug-related issue reports is the same as observed in BugZilla, results are relatively close to the ones resulting from models trained and tested with BugZilla projects. Table \ref{tab_its_complementary} demonstrates with some examples how the F-measure of the complementary models varies due to class imbalance in the testing set. In the table, the columns \textit{Mean Prop. Bugs} and \textit{Mean Prop. Non-bugs} represent the mean proportion of bug and non-bug-related issue reports in the testing set, respectively. The complementary experiments we run suggest the performance variations observed in RQ4 are associated with differences in the prevalence of classes observed in the datasets.









\begin{table}[!htp]\centering
\caption{F-score variation of Jira C/C++ models with different class proportions in the testing set.}
\label{tab_its_complementary}
\scriptsize
\begin{tabular}{lrrrr}\toprule
\textbf{ITS} &\textbf{Mean Prop. Bugs} &\textbf{Mean Prop. Non-bugs} &\textbf{Mean F-score} \\\midrule
Jira &78\% &22\% &0.56 \\
Jira &70\% &30\% &0.65 \\
Jira &60\% &40\% &0.72 \\
Jira &50\% &50\% &0.77 \\
\bottomrule
\end{tabular}
\end{table}

\section{Cross-Project Classification (RQ5)} 
\label{section_res_rq5}

The main goal of RQ5 is to understand how well ML algorithms can perform automatic classification of issue reports coming for projects that were not present during the training phase, which we refer to as \textit{cross-project classification}. We selected five projects written in the same language (i.e., Java) and that rely on the same ITS (i.e., GitHub), namely Bazel, Elasticsearch, Netty, Spring Boot and Spring Framework, as previously detailed in Table \ref{tab_projects}. Our intention with this selection of projects is to eliminate possible variations in the F-measure that might occur due to programming language and ITS, as we have previously observed that these are factors that influence results (see RQ3 and RQ4). Again, we present the results obtained with the SVM algorithm.

\begin{figure}[!h]
\centerline{\includegraphics[width=65mm]{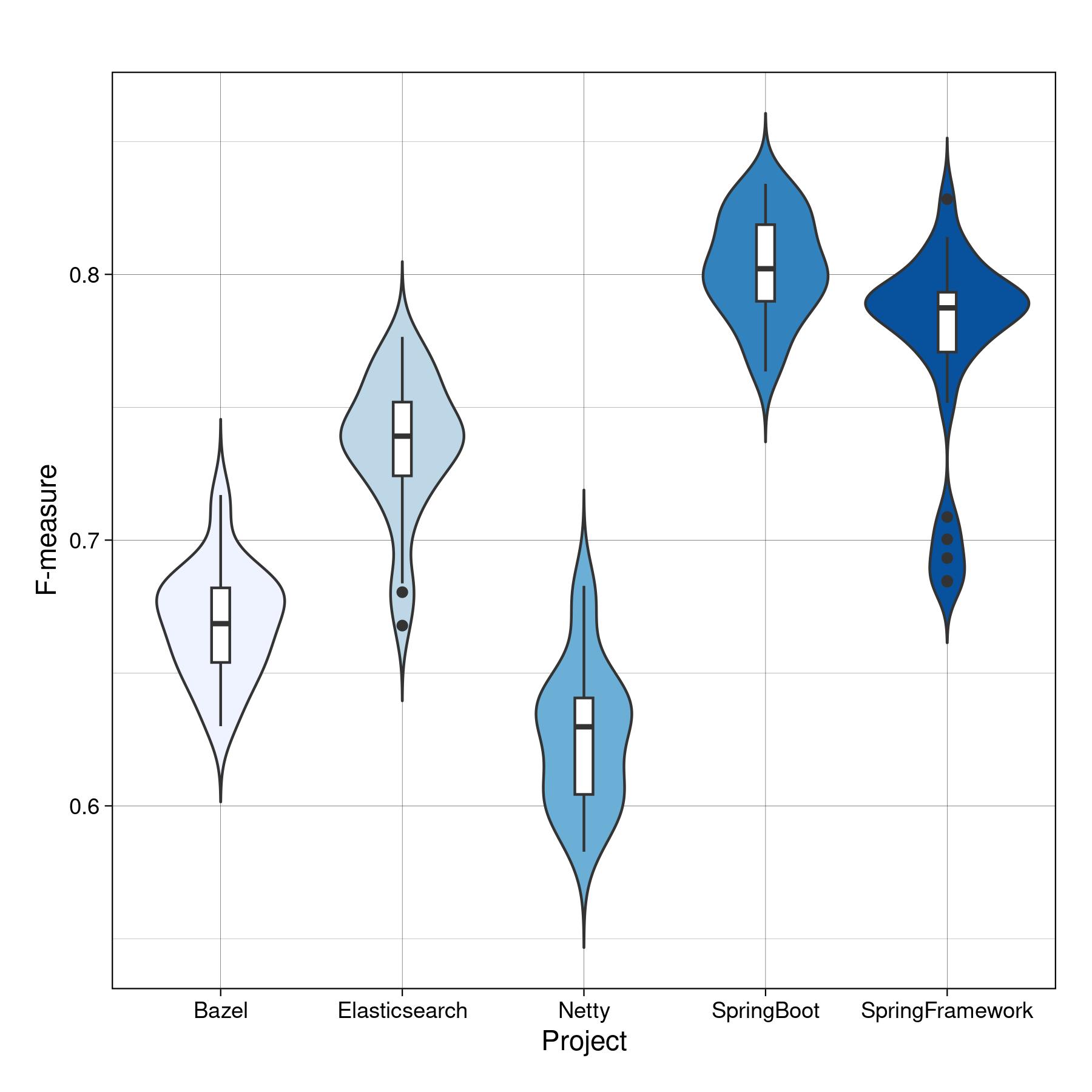}}
\caption{Results for training and testing with the same project.}
\label{vl_projects_sp}
\end{figure}

In order to have a baseline reference, prior to the \textit{cross-project classification}, we trained and tested models with issue reports from the same software system. We use the same amount of issue reports for each of the 5 projects to avoid having the results biased towards the projects that have more issue reports. Thus, for each project, we randomly selected 2,000 reports, which we divided in the usual proportion for training and testing (i.e., 1400 reports for training, 700 for testing). As in the previous RQs, we ensured that the samples for training are balanced (50\% represent bugs and the remaining represent non-bugs).
Figure \ref{vl_projects_sp} presents the F-measure results obtained. It visually indicates differences between all projects. We then verified if these differences were statistically significant, which was indeed the case. This indicates that, \textit{even for projects written in the same language and supported by the same ITS we observe significant differences in the classification effectiveness}. 


\begin{figure}[!h]
\centerline{\includegraphics[width=65mm]{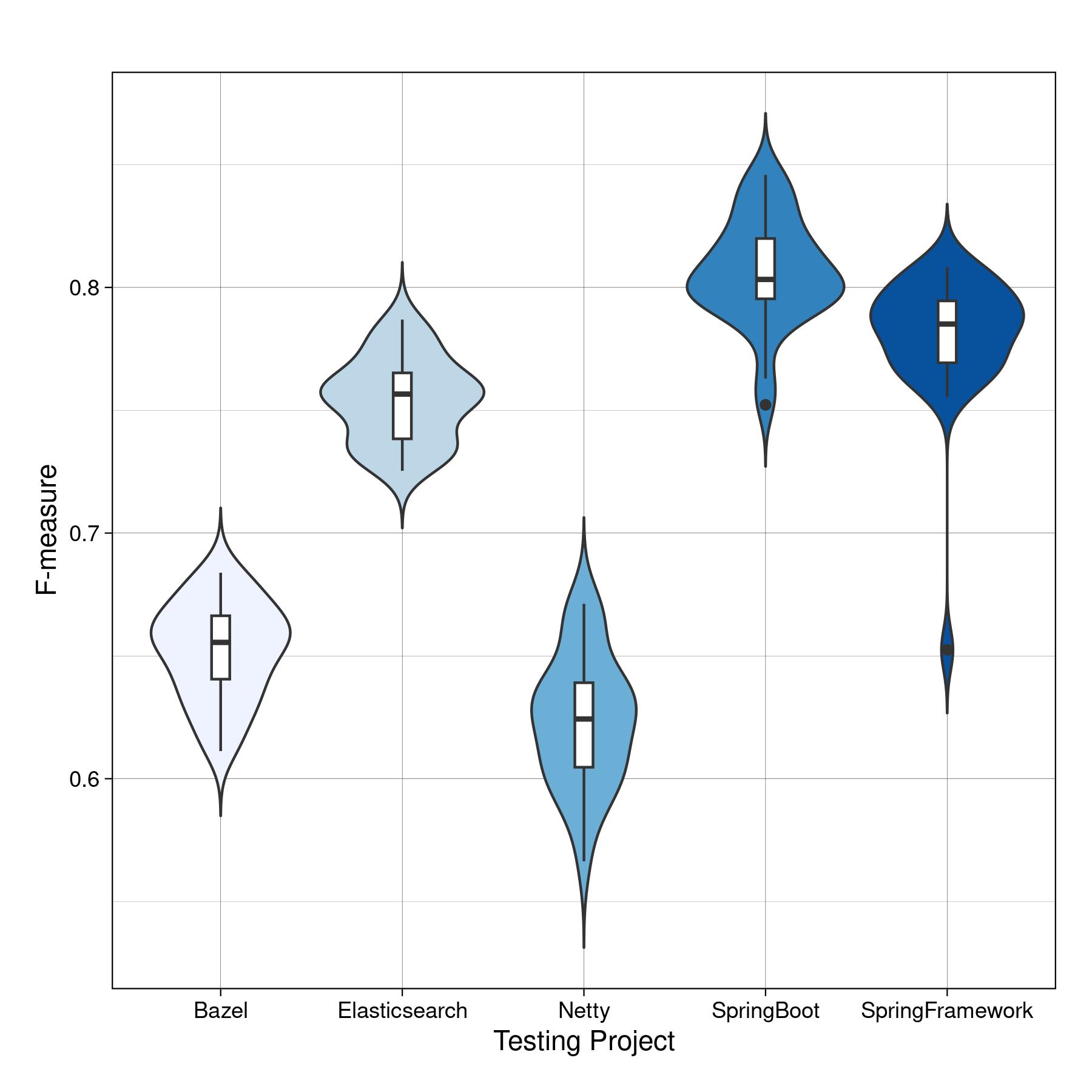}}
\caption{Cross-project classification with SVM.}
\label{vl_projects_dp}
\end{figure}

To create models for \textit{cross project classification}, we trained SVM models with four of the five selected projects and tested them with the remaining one. As we are using five projects, this is done five times, to have results for each of the projects participating in the tests. The process begins with the random collection of a total of 6,000 samples from the four projects (i.e., 1,500 issue reports of each project, in a proportion of 50\% bugs and 50\% non-bugs). From the testing project (i.e., the fifth project, different from the ones used for training), we randomly select 600 samples, maintaining the original proportion of bugs and non-bugs. Thus, this process is repeated in a round-robin manner, until all projects pass through the testing phase and all respective metrics are collected. Figure \ref{vl_projects_dp} presents the F-measure values obtained with SVM for each of the tested projects.

The projects in Figure \ref{vl_projects_dp} show visible differences among themselves. The average F-measure values achieved by SVM for Elasticsearch, Netty, Bazel, SpringBoot and SpringFramework are respectively 0.76, 0.62, 0.66, 0.81 and 0.78, and the interquartile ranges and distributions are also visually different. We statistically tested whether the visually observed differences in F-measure values were significant depending on the testing project, with the results confirming that indeed \textit{there are significant differences depending on the project being tested} 

Figures \ref{vl_projects_sp} and \ref{vl_projects_dp} show relatively similar F-measure values and, in this case, it is difficult to distinguish clear visual differences between the pairs of projects in both figures. Thus, we statistically tested if training/testing within the same project leads to significantly different results from when training using a mix of projects and testing in a new one. We found out \textit{p-values} higher than the significance level of \(\alpha=0.05\) for three projects (i.e., Netty, Spring Framework and Spring Boot). On the other hand, we also discovered two cases in which the \textit{p-values} were lower than the significance level, namely for Elasticsearch and Bazel models. Based on such findings, we reached the following conclusion:

\begin{tcolorbox}[left=1pt,top=1pt,bottom=1pt,right=1pt,colback=gray!2!white,colframe=gray!75!gray!50!]
    \textbf{Response to RQ5:} The performance of SVM models trained and tested with data from the same project compared with the performance of SVM models tested with data from projects not present during training may not always be significant (considering projects based on the same tracking system and programming language).
\end{tcolorbox}

As in the previous research questions, we observed consistent results across all algorithms used in the paper. We also conducted further analyses to understand the reasons behind our findings. We analyzed the TF-IDF terms selected by the chi-squared method, which are shared among different projects written in the same programming language, and using the same ITS. Our analysis reveals that only just a few terms are relevant between two projects simultaneously. For instance, the amount of shared terms between any two projects is on average 9\%. This means that in the experiments we made in the scope of our fifth RQ, around 36\% of the terms appear in both the training and testing sets. Our analysis indicates that increasing the number of projects used for training could be a solution for creating more effective models for cross-project classification, as long as some aspects of the projects are respected, such as programming languages and issue tracking systems. Our findings contribute to answering the open question in the literature precisely related to cross-project and cross-ITSs classification. Also, this outcome will help in future research, when the goal is to use automatic classification for new projects or simply projects that do not have bug reports in sufficient quantity.
\section{Guidelines for Future Studies} 
\label{sec_guidelines}

Based on the results discussed above, we compiled a set of guidelines for future studies on the automatic classification of issue reports using ML algorithms. In summary, the following should be taken into consideration:
\begin{enumerate}
    \item \textit{Data acquisition}: data should be as heterogeneous as possible. First, we recommend using projects of different natures. Also, different programming languages and tracking systems shall be considered as criteria to select the projects. Choosing projects that have a minimum amount of issue reports is equally important.
    \item \textit{Data preprocessing}: our experiments suggest that stop-word removal should be used with caution, as removing some so-called \textit{stop-words} can alter the meaning of important sentences. To avoid such problems, our recommendation is to carefully revise the list of stop-words. In addition, our results suggest that using lemmatisation is preferable over stemming. Removing punctuation and converting sentences to lowercase can also be useful.
    \item \textit{Feature engineering}: bag-of-words has proven to be reasonable for the automatic classification of issue reports. Yet, TF-IDF should be used, instead of only TF, as it gives stronger weights to terms that are in fact important to the decision process. Also, chi-squered feature selection method has shown itself as a good tradeoff between efficiency and required processing power.
    \item \textit{Model selection}: our experiments suggest that SVM, RF and LR should be adopted instead of NB and KNN classifiers, as we demonstrate that the former achieve superior performance.
    \item \textit{Training models}: we suggest randomizing and balancing the training data. This is important as the experiments should be repeated several times to reduce bias and increase reliance on the results, and it does not make sense to repeat experiments in which no randomization is observed. We recommend using a GridSearch and carrying out a hyperparameter optimization for each training, if possible. Yet, as it requires reasonable processing power, a set of hyperparameter optimization can be run by using a representative sample of the data, to find the best suitable parameters to use for each classifier.
    \item \textit{Model evaluation}: we recommend testing models with imbalanced data, in the original proportion as collected from the ITS. For results assessment, we suggest the adoption of F-measure, instead of accuracy, as the last can lead to questionable results when imbalanced data is being used. Also, looking at \textit{precision} and \textit{recall} could be useful to increase reliance on the results evaluation procedure. Finally, less traditional metrics such as ROC and AUC can be helpful whenever they can be used.
\end{enumerate}
\section{Threats to Validity} 
\label{sec_thereats}


This section addresses the main threats to the validity of our study and explains how we mitigate them.

Regarding \textbf{internal validity}, it is relevant to point out that  \textit{the selection of projects may add bias to the results}. To minimize this issue, we tried to use a broad set of projects encompassing different types of software systems, with different sizes and complexity, written in various programming languages and supported by distinct issue tracking systems. In the end, just in terms of size, we are already using a dataset that is larger what is generally observed in the state of the art.

Regarding the overall algorithm configuration, it is relevant to mention that the \textit{hyperparameter configuration may not be optimal and may influence the overall results}. Indeed, there is always a balance to reach between the search space exploration and the computational effort required to find the best parameters. In this study it would be not feasible to perform such a task for each execution, considering that we executed 30 repetitions and that we carried out more than 785 experiments. The solution we came across was to execute the optimization several times with random samples extracted from different projects, in order to obtain a set of parameters that would be generally suitable, which we find to be a reasonable solution that is often used in similar research. 

Certain choices, such as \textit{the selection of feature values in increments of 50 units (e.g., 50, 100, 150, 200, ...) may have effect on the results}. In this particular case, we had observed works using 50, 100, and 150 features, thus we maintained this setting and explored larger values, using 50 units as the base interval for the experiments. Another example is the fact that \textit{the option for Bag of Words and TF-IDF may have some impact on the results}. Our choice is related with the nature of this work, which primarily aims at classic, widely used methods (e.g., TF-IDF is quite popular in this context, as we observed in related work), in a way to build a base for future comparison with other techniques.

\textit{The selection of data samples for training and testing may add bias to the models' performance}, with some samples becoming over- or under-represented. To avoid this issue, in each train/test sequence, we used a number of issue reports that were lower than the total amount of samples available (to allow for selecting different samples in different runs). We also used random sampling and shuffled the data before selecting the examples. 
To mitigate possible \textit{variability in the model's performance due to randomness in the data splitting process}, we also ran each train/test execution 30 times.

The \textit{data used in this work may not be fully reliable}. We targeted only "resolved" and "fixed" issue reports, which means that they have been previously analyzed by developers and should be correctly labeled. We verified our acquisition tool (and also checked the result of its operation by randomly looking at the samples extracted). Any existing errors in the data should be residual, if present at all. It is also important to mention that \textit{bad quality data may exist in the dataset}, for which we used common preprocessing techniques to allow models to extract meaningful patterns more easily. 

Concerning \textbf{external validity}, a common problem is the \textit{curse of dimensionality}, in which the
number of features or dimensions is large relative to the number of samples. In practice, the effect is that the models overfit to the specific set of data used during training and cannot really generalize.
To mitigate this issue, we performed dimensionality reduction using the chi-squared method \cite{Novakovic2011}
, due to the fact that it was successfully used by other authors in a context similar to our own.

The \textit{data used is not be representative of all software projects}. This may affect the generalization of the results but we aimed at a large and heterogeneous set of projects, which, to the best of our knowledge, ended up in the largest dataset of its kind.

To ensure \textit{the reproducibility of results} we mare making our datasets publicly available, as well as the tools we developed to automate tasks and the scripts we created within the scope of this study at \cite{SupMaterials2022}.

\section{Conclusion} 
\label{sec_conclusion}
In this paper, we analyzed the effectiveness of several ML algorithms in the classification of bug reports, considering whether a particular issue report refers to a real bug or not. We observed that when using BoW, 250 dimensions are adequate to classify bug reports and that using either the report title or the description does not make a difference in the classification effectiveness. 
We also found out that many classifiers achieve comparable results, while NB and KNN, which are commonly-used algorithms in this context, are associated with poor performance. 

During the experimental evaluation, we observed differences in classification effectiveness associated with projects written in different programming languages and also with projects supported by different ITS. 
Also, we noticed differences between the performance of models trained and tested with data from the same project and the performance of models tested with issue reports from systems not used in the training process. However, as long as the same issue tracking system and programming language are respected and representative data are used for training, the models generated from bug reports of heterogeneous sources are sufficiently robust and can be used to carry out cross-project classification effectively. Finally, we observed that, when performing cross-project classification, the performance of models may vary more due to testing examples than the training data itself, as long as high-quality issue reports are used to train the classifiers.

The analysis of our results, allowed us to design a series of guidelines for further research in bug report classification. Overall, we highlight the fact that the use of heterogeneous data (i.e., different projects, programming languages, issue tracking systems) is important to allow models to learn from the various facets involved and to create models that are effective in classification. Characteristics or differences in the data that may influence classification results should be taken into account when designing new studies. For instance, if such studies aim at showing the classification effectiveness of different algorithms, possible influencing factors should be disclosed (e.g., programming language).



In future work, we intend to specifically target the interpretability of the classification models by further analyzing the features that most strongly influence the classification outcomes. Also, we intend to study other forms of data as input for the classification, such as code excerpts, characteristics of developers, or nature of the systems involved. Our intention is also to further explore additional techniques in the classification pipeline, such as Principal Component Analysis or Latent Semantic Analysis, as well as embedding models such as BERT, CodeBERT, or CodeT5, just to name a few. 

We plan to further analyze the ability of the algorithms in multi-class problems, namely in what concerns the types of bugs involved, the severity of the reports and automatic handover of reports for developers. We also intend to study how other classes of algorithms (e.g., deep learning and large language models) can be used to perform issue report classification tasks.



\section*{Acknowledgments}

This work has been partially supported by the FCT – Foundation for Science and Technology, I.P./MCTES through national funds (PIDDAC), within the scope of CISUC R\&D Unit – UIDB/00326/2020 or project code UIDP/00326/2020; and by Project “NEXUS Pacto de Inovação – Transição Verde e Digital para Transportes, Logística e Mobilidade”. ref. No. 7113, supported by the Recovery and Resilience Plan (PRR) and by the European Funds Next Generation EU, following Notice No. 02/C05-i01/2022.PC645112083-00000059 (project 53), Component 5 - Capitalization and Business Innovation - Mobilizing Agendas for Business Innovation; and by the Portuguese Foundation for Science and Technology (FCT), through the PhD grant UI/BD/151217/2021, within the scope of the project CISUC - UID/CEC/00326/2020 and by the European Social Fund, through the Regional Operational Program Centro 2020; and by INCD funded by FCT and FEDER under project 01/SAICT/2016 nº 022153.

\bibliographystyle{ACM-Reference-Format}
\bibliography{references}


\begin{thebibliography}{69}


\ifx \showCODEN    \undefined \def \showCODEN     #1{\unskip}     \fi
\ifx \showDOI      \undefined \def \showDOI       #1{#1}\fi
\ifx \showISBNx    \undefined \def \showISBNx     #1{\unskip}     \fi
\ifx \showISBNxiii \undefined \def \showISBNxiii  #1{\unskip}     \fi
\ifx \showISSN     \undefined \def \showISSN      #1{\unskip}     \fi
\ifx \showLCCN     \undefined \def \showLCCN      #1{\unskip}     \fi
\ifx \shownote     \undefined \def \shownote      #1{#1}          \fi
\ifx \showarticletitle \undefined \def \showarticletitle #1{#1}   \fi
\ifx \showURL      \undefined \def \showURL       {\relax}        \fi
\providecommand\bibfield[2]{#2}
\providecommand\bibinfo[2]{#2}
\providecommand\natexlab[1]{#1}
\providecommand\showeprint[2][]{arXiv:#2}

\bibitem[Abhishek and Abdelaziz(2023)]%
        {abhishek2023}
\bibfield{author}{\bibinfo{person}{K Abhishek} {and} \bibinfo{person}{M Abdelaziz}.} \bibinfo{year}{2023}\natexlab{}.
\newblock \bibinfo{booktitle}{\emph{{Machine Learning for Imbalanced Data: Tackle imbalanced datasets using machine learning and deep learning techniques}}}.
\newblock \bibinfo{publisher}{Packt Publishing}.
\newblock
\showISBNx{9781801070881}
\urldef\tempurl%
\url{https://books.google.pt/books?id=xx_nEAAAQBAJ}
\showURL{%
\tempurl}


\bibitem[Aggarwal(2015)]%
        {Aggarwal2015}
\bibfield{author}{\bibinfo{person}{Charu~C Aggarwal}.} \bibinfo{year}{2015}\natexlab{}.
\newblock \bibinfo{booktitle}{\emph{{Data Mining: The Textbook}}}.
\newblock \bibinfo{publisher}{Springer Publishing Company, Incorporated}.
\newblock
\showISBNx{978-3-319-14142-8}
\urldef\tempurl%
\url{https://doi.org/10.1007/978-3-319-14142-8}
\showDOI{\tempurl}


\bibitem[Andrade et~al\mbox{.}(2024)]%
        {Andrade2024}
\bibfield{author}{\bibinfo{person}{Renato Andrade}, \bibinfo{person}{Nuno Laranjeiro}, {and} \bibinfo{person}{Marco Vieira}.} \bibinfo{year}{2024}\natexlab{}.
\newblock \showarticletitle{{BugHub: A Large Scale Issue Report Dataset}}. In \bibinfo{booktitle}{\emph{19th European Dependable Computing Conference (EDCC), accepted for publication}}. \bibinfo{address}{Leuven, Belgium}.
\newblock
\urldef\tempurl%
\url{https://eden.dei.uc.pt/~renatoandrade/_2023_EDCC__Bug_Report_Dataset.pdf}
\showURL{%
\tempurl}


\bibitem[Andrade et~al\mbox{.}(2025)]%
        {SupMaterials2022}
\bibfield{author}{\bibinfo{person}{Renato Andrade}, \bibinfo{person}{César Teixeira}, \bibinfo{person}{Nuno Laranjeiro}, {and} \bibinfo{person}{Marco Vieira}.} \bibinfo{year}{2025}\natexlab{}.
\newblock \bibinfo{title}{{An Empirical Study on the Classification of Bug Reports Using Machine Learning Algorithms - Supplementary Material}}.
\newblock
\newblock
\urldef\tempurl%
\url{https://doi.org/10.5281/zenodo.7377401}
\showDOI{\tempurl}


\bibitem[Antoniol et~al\mbox{.}(2008)]%
        {Antoniol2008}
\bibfield{author}{\bibinfo{person}{Giuliano Antoniol}, \bibinfo{person}{Kamel Ayari}, \bibinfo{person}{Massimiliano {Di Penta}}, \bibinfo{person}{Foutse Khomh}, {and} \bibinfo{person}{Yann~Ga{\"{e}}l Gu{\'{e}}h{\'{e}}neuc}.} \bibinfo{year}{2008}\natexlab{}.
\newblock \showarticletitle{{Is it a bug or an enhancement? A text-based approach to classify change requests}}.
\newblock \bibinfo{journal}{\emph{Proceedings of the 2008 Conference of the Center for Advanced Studies, CASCON'08}} (\bibinfo{year}{2008}).
\newblock
\urldef\tempurl%
\url{https://doi.org/10.1145/1463788.1463819}
\showDOI{\tempurl}


\bibitem[Azeem et~al\mbox{.}(2019)]%
        {Azeem2019}
\bibfield{author}{\bibinfo{person}{Muhammad~Ilyas Azeem}, \bibinfo{person}{Fabio Palomba}, \bibinfo{person}{Lin Shi}, {and} \bibinfo{person}{Qing Wang}.} \bibinfo{year}{2019}\natexlab{}.
\newblock \showarticletitle{Machine learning techniques for code smell detection: A systematic literature review and meta-analysis}.
\newblock \bibinfo{journal}{\emph{Information and Software Technology}}  \bibinfo{volume}{108} (\bibinfo{year}{2019}), \bibinfo{pages}{115--138}.
\newblock
Issue 4.
\showISSN{09505849}
\urldef\tempurl%
\url{https://doi.org/10.1016/j.infsof.2018.12.009}
\showDOI{\tempurl}


\bibitem[Babanejad et~al\mbox{.}(2020)]%
        {Babanejad2020}
\bibfield{author}{\bibinfo{person}{Nastaran Babanejad}, \bibinfo{person}{Ameeta Agrawal}, \bibinfo{person}{Aijun An}, {and} \bibinfo{person}{Manos Papagelis}.} \bibinfo{year}{2020}\natexlab{}.
\newblock \showarticletitle{{A Comprehensive Analysis of Preprocessing for Word Representation Learning in Affective Tasks}}. In \bibinfo{booktitle}{\emph{Proceedings of the 58th Annual Meeting of the Association for Computational Linguistics}}. \bibinfo{publisher}{Association for Computational Linguistics}, \bibinfo{address}{Online}, \bibinfo{pages}{5799--5810}.
\newblock
\urldef\tempurl%
\url{https://doi.org/10.18653/v1/2020.acl-main.514}
\showDOI{\tempurl}


\bibitem[Breiman(2001)]%
        {Breiman2001}
\bibfield{author}{\bibinfo{person}{Leo Breiman}.} \bibinfo{year}{2001}\natexlab{}.
\newblock \showarticletitle{{Random Forests}}.
\newblock \bibinfo{journal}{\emph{Machine Learning}} \bibinfo{volume}{45}, \bibinfo{number}{1} (\bibinfo{year}{2001}), \bibinfo{pages}{5--32}.
\newblock
\showISSN{1573-0565}
\urldef\tempurl%
\url{https://doi.org/10.1023/A:1010933404324}
\showDOI{\tempurl}


\bibitem[Catolino et~al\mbox{.}(2019)]%
        {Catolino2019}
\bibfield{author}{\bibinfo{person}{Gemma Catolino}, \bibinfo{person}{Fabio Palomba}, \bibinfo{person}{Andy Zaidman}, {and} \bibinfo{person}{Filomena Ferrucci}.} \bibinfo{year}{2019}\natexlab{}.
\newblock \showarticletitle{Not all bugs are the same: Understanding, characterizing, and classifying bug types}.
\newblock \bibinfo{journal}{\emph{Journal of Systems and Software}}  \bibinfo{volume}{152} (\bibinfo{year}{2019}), \bibinfo{pages}{165--181}.
\newblock
\showISSN{01641212}
\urldef\tempurl%
\url{https://doi.org/10.1016/j.jss.2019.03.002}
\showDOI{\tempurl}


\bibitem[Chawla and Singh(2015)]%
        {Chawla2015}
\bibfield{author}{\bibinfo{person}{Indu Chawla} {and} \bibinfo{person}{Sandeep~K. Singh}.} \bibinfo{year}{2015}\natexlab{}.
\newblock \showarticletitle{An automated approach for bug categorization using fuzzy logic}.
\newblock \bibinfo{journal}{\emph{ACM International Conference Proceeding Series}}  \bibinfo{volume}{18-20-Febr} (\bibinfo{year}{2015}), \bibinfo{pages}{90--99}.
\newblock
\showISBNx{9781450334327}
\urldef\tempurl%
\url{https://doi.org/10.1145/2723742.2723751}
\showDOI{\tempurl}


\bibitem[Cortes and Vapnik(1992)]%
        {Cortes1992}
\bibfield{author}{\bibinfo{person}{Corinna Cortes} {and} \bibinfo{person}{Vladimir Vapnik}.} \bibinfo{year}{1992}\natexlab{}.
\newblock \showarticletitle{{Support-Vector Networks}}.
\newblock \bibinfo{journal}{\emph{IEEE Expert-Intelligent Systems and their Applications}} \bibinfo{volume}{7}, \bibinfo{number}{5} (\bibinfo{year}{1992}), \bibinfo{pages}{63--72}.
\newblock
\showISSN{08859000}
\urldef\tempurl%
\url{https://doi.org/10.1109/64.163674}
\showDOI{\tempurl}


\bibitem[Cunningham and Delany(2021)]%
        {Cunningham2021}
\bibfield{author}{\bibinfo{person}{P{\'{a}}draig Cunningham} {and} \bibinfo{person}{Sarah~Jane Delany}.} \bibinfo{year}{2021}\natexlab{}.
\newblock \showarticletitle{{K-Nearest Neighbour Classifiers - A Tutorial}}.
\newblock \bibinfo{journal}{\emph{ACM Comput. Surv.}} \bibinfo{volume}{54}, \bibinfo{number}{6} (\bibinfo{date}{jul} \bibinfo{year}{2021}).
\newblock
\showISSN{0360-0300}
\urldef\tempurl%
\url{https://doi.org/10.1145/3459665}
\showDOI{\tempurl}


\bibitem[Du et~al\mbox{.}(2017)]%
        {Du2017}
\bibfield{author}{\bibinfo{person}{Xiaoting Du}, \bibinfo{person}{Zheng Zheng}, \bibinfo{person}{Guanping Xiao}, {and} \bibinfo{person}{Beibei Yin}.} \bibinfo{year}{2017}\natexlab{}.
\newblock \showarticletitle{{The automatic classification of fault trigger based bug report}}.
\newblock \bibinfo{journal}{\emph{Proceedings - 2017 IEEE 28th International Symposium on Software Reliability Engineering Workshops, ISSREW 2017}} (\bibinfo{year}{2017}), \bibinfo{pages}{259--265}.
\newblock
\showISBNx{9781538623879}
\urldef\tempurl%
\url{https://doi.org/10.1109/ISSREW.2017.28}
\showDOI{\tempurl}


\bibitem[Fan et~al\mbox{.}(2017)]%
        {Fan2017}
\bibfield{author}{\bibinfo{person}{Qiang Fan}, \bibinfo{person}{Yue Yu}, \bibinfo{person}{Gang Yin}, \bibinfo{person}{Tao Wang}, {and} \bibinfo{person}{Huaimin Wang}.} \bibinfo{year}{2017}\natexlab{}.
\newblock \showarticletitle{Where Is the Road for Issue Reports Classification Based on Text Mining?}
\newblock \bibinfo{journal}{\emph{International Symposium on Empirical Software Engineering and Measurement}}  \bibinfo{volume}{2017-Novem} (\bibinfo{year}{2017}), \bibinfo{pages}{121--130}.
\newblock
\showISBNx{9781509040391}
\showISSN{19493789}
\urldef\tempurl%
\url{https://doi.org/10.1109/ESEM.2017.19}
\showDOI{\tempurl}


\bibitem[Fan et~al\mbox{.}(2020)]%
        {Fan2020}
\bibfield{author}{\bibinfo{person}{Yuanrui Fan}, \bibinfo{person}{Xin Xia}, \bibinfo{person}{David Lo}, {and} \bibinfo{person}{Ahmed~E. Hassan}.} \bibinfo{year}{2020}\natexlab{}.
\newblock \showarticletitle{{Chaff from the Wheat: Characterizing and Determining Valid Bug Reports}}.
\newblock \bibinfo{journal}{\emph{IEEE Transactions on Software Engineering}} \bibinfo{volume}{46}, \bibinfo{number}{5} (\bibinfo{year}{2020}), \bibinfo{pages}{495--525}.
\newblock
\showISSN{19393520}
\urldef\tempurl%
\url{https://doi.org/10.1109/TSE.2018.2864217}
\showDOI{\tempurl}


\bibitem[Fern{\'a}ndez et~al\mbox{.}(2018)]%
        {fernández2018}
\bibfield{author}{\bibinfo{person}{A. Fern{\'a}ndez}, \bibinfo{person}{S. Garc{\'\i}a}, \bibinfo{person}{M. Galar}, \bibinfo{person}{R.C. Prati}, \bibinfo{person}{B. Krawczyk}, {and} \bibinfo{person}{F. Herrera}.} \bibinfo{year}{2018}\natexlab{}.
\newblock \bibinfo{booktitle}{\emph{{Learning from Imbalanced Data Sets}}}.
\newblock \bibinfo{publisher}{Springer International Publishing}.
\newblock
\showISBNx{9783319980744}
\urldef\tempurl%
\url{https://books.google.pt/books?id=8Fp0DwAAQBAJ}
\showURL{%
\tempurl}


\bibitem[Fligner and Killeen(1976)]%
        {Fligner1976}
\bibfield{author}{\bibinfo{person}{Michael~A Fligner} {and} \bibinfo{person}{Timothy~J Killeen}.} \bibinfo{year}{1976}\natexlab{}.
\newblock \showarticletitle{{Distribution-Free Two-Sample Tests for Scale}}.
\newblock \bibinfo{journal}{\emph{J. Amer. Statist. Assoc.}} \bibinfo{volume}{71}, \bibinfo{number}{353} (\bibinfo{year}{1976}), \bibinfo{pages}{210--213}.
\newblock
\showISSN{01621459}
\urldef\tempurl%
\url{https://doi.org/10.2307/2285771}
\showDOI{\tempurl}


\bibitem[Fontana and Zanoni(2017)]%
        {ArcelliFontana2017}
\bibfield{author}{\bibinfo{person}{Francesca~Arcelli Fontana} {and} \bibinfo{person}{Marco Zanoni}.} \bibinfo{year}{2017}\natexlab{}.
\newblock \showarticletitle{Code smell severity classification using machine learning techniques}.
\newblock \bibinfo{journal}{\emph{Knowledge-Based Systems}}  \bibinfo{volume}{128} (\bibinfo{year}{2017}), \bibinfo{pages}{43--58}.
\newblock
\showISSN{09507051}
\urldef\tempurl%
\url{https://doi.org/10.1016/j.knosys.2017.04.014}
\showDOI{\tempurl}


\bibitem[Hapke and Nelson(2020)]%
        {hapke2020}
\bibfield{author}{\bibinfo{person}{H Hapke} {and} \bibinfo{person}{C Nelson}.} \bibinfo{year}{2020}\natexlab{}.
\newblock \bibinfo{booktitle}{\emph{Building Machine Learning Pipelines}}.
\newblock \bibinfo{publisher}{O'Reilly Media}.
\newblock
\showISBNx{9781492053149}


\bibitem[He and Ma(2013)]%
        {he2013}
\bibfield{author}{\bibinfo{person}{H He} {and} \bibinfo{person}{Y Ma}.} \bibinfo{year}{2013}\natexlab{}.
\newblock \bibinfo{booktitle}{\emph{{Imbalanced Learning: Foundations, Algorithms, and Applications}}}.
\newblock \bibinfo{publisher}{Wiley}.
\newblock
\showISBNx{9781118646335}
\urldef\tempurl%
\url{https://books.google.pt/books?id=CVHx-Gp9jzUC}
\showURL{%
\tempurl}


\bibitem[He et~al\mbox{.}(2020)]%
        {He2020}
\bibfield{author}{\bibinfo{person}{Jianjun He}, \bibinfo{person}{Ling Xu}, \bibinfo{person}{Meng Yan}, \bibinfo{person}{Xin Xia}, {and} \bibinfo{person}{Yan Lei}.} \bibinfo{year}{2020}\natexlab{}.
\newblock \showarticletitle{{Duplicate bug report detection using dual-channel convolutional neural networks}}.
\newblock \bibinfo{journal}{\emph{IEEE International Conference on Program Comprehension}} (\bibinfo{year}{2020}), \bibinfo{pages}{117--127}.
\newblock
\showISBNx{9781450379588}
\urldef\tempurl%
\url{https://doi.org/10.1145/3387904.3389263}
\showDOI{\tempurl}


\bibitem[Herzig et~al\mbox{.}(2013)]%
        {Herzig2013}
\bibfield{author}{\bibinfo{person}{Kim Herzig}, \bibinfo{person}{Sascha Just}, {and} \bibinfo{person}{Andreas Zeller}.} \bibinfo{year}{2013}\natexlab{}.
\newblock \showarticletitle{{It's not a bug, it's a feature: How misclassification impacts bug prediction}}.
\newblock \bibinfo{journal}{\emph{Proceedings - International Conference on Software Engineering}} (\bibinfo{year}{2013}), \bibinfo{pages}{392--401}.
\newblock
\showISBNx{9781467330763}
\showISSN{02705257}
\urldef\tempurl%
\url{https://doi.org/10.1109/ICSE.2013.6606585}
\showDOI{\tempurl}


\bibitem[Hsu(1996)]%
        {hsu1996}
\bibfield{author}{\bibinfo{person}{Jason~C Hsu}.} \bibinfo{year}{1996}\natexlab{}.
\newblock \bibinfo{booktitle}{\emph{Multiple Comparisons: Theory and methods}}.
\newblock \bibinfo{publisher}{Springer US}.
\newblock
\showISBNx{9780412982811; 0412982811; 9781489971807; 1489971807}


\bibitem[Japkowicz and Shah(2011)]%
        {Japkowicz2011}
\bibfield{author}{\bibinfo{person}{N Japkowicz} {and} \bibinfo{person}{M Shah}.} \bibinfo{year}{2011}\natexlab{}.
\newblock \bibinfo{booktitle}{\emph{{Evaluating Learning Algorithms: A Classification Perspective}}}.
\newblock \bibinfo{publisher}{Cambridge University Press}.
\newblock
\showISBNx{9781139494144}
\urldef\tempurl%
\url{https://books.google.com.br/books?id=VoWIIOKVzR4C}
\showURL{%
\tempurl}


\bibitem[Kallis et~al\mbox{.}(2019)]%
        {Kallis2019}
\bibfield{author}{\bibinfo{person}{Rafael Kallis}, \bibinfo{person}{Andrea~Di Sorbo}, \bibinfo{person}{Gerardo Canfora}, {and} \bibinfo{person}{Sebastiano Panichella}.} \bibinfo{year}{2019}\natexlab{}.
\newblock \showarticletitle{Ticket Tagger: Machine Learning Driven Issue Classification}.
\newblock \bibinfo{journal}{\emph{Proceedings - 2019 IEEE International Conference on Software Maintenance and Evolution, ICSME 2019}} (\bibinfo{year}{2019}), \bibinfo{pages}{406--409}.
\newblock
\showISBNx{9781728130941}
\urldef\tempurl%
\url{https://doi.org/10.1109/ICSME.2019.00070}
\showDOI{\tempurl}


\bibitem[Ko et~al\mbox{.}(2006)]%
        {ko2006}
\bibfield{author}{\bibinfo{person}{Andrew~J Ko}, \bibinfo{person}{Brad~A Myers}, {and} \bibinfo{person}{Duen~Horng Chau}.} \bibinfo{year}{2006}\natexlab{}.
\newblock \showarticletitle{A linguistic analysis of how people describe software problems}. In \bibinfo{booktitle}{\emph{Visual Languages and Human-Centric Computing (VL/HCC'06)}}. IEEE, \bibinfo{pages}{127--134}.
\newblock


\bibitem[Kochhar et~al\mbox{.}(2014)]%
        {Kochhar2014}
\bibfield{author}{\bibinfo{person}{Pavneet~Singh Kochhar}, \bibinfo{person}{Ferdian Thung}, {and} \bibinfo{person}{David Lo}.} \bibinfo{year}{2014}\natexlab{}.
\newblock \showarticletitle{Automatic fine-grained issue report reclassification}.
\newblock \bibinfo{journal}{\emph{Proceedings of the IEEE International Conference on Engineering of Complex Computer Systems, ICECCS}} (\bibinfo{year}{2014}), \bibinfo{pages}{126--135}.
\newblock
\showISBNx{9781479954827}
\urldef\tempurl%
\url{https://doi.org/10.1109/ICECCS.2014.25}
\showDOI{\tempurl}


\bibitem[Korenius et~al\mbox{.}(2004)]%
        {Korenius2004}
\bibfield{author}{\bibinfo{person}{Tuomo Korenius}, \bibinfo{person}{Jorma Laurikkala}, \bibinfo{person}{Kalervo J\"{a}rvelin}, {and} \bibinfo{person}{Martti Juhola}.} \bibinfo{year}{2004}\natexlab{}.
\newblock \showarticletitle{Stemming and Lemmatization in the Clustering of Finnish Text Documents}. In \bibinfo{booktitle}{\emph{Proceedings of the Thirteenth ACM International Conference on Information and Knowledge Management}} (Washington, D.C., USA) \emph{(\bibinfo{series}{CIKM '04})}. \bibinfo{publisher}{Association for Computing Machinery}, \bibinfo{address}{New York, NY, USA}, \bibinfo{pages}{625–633}.
\newblock
\showISBNx{1581138741}
\urldef\tempurl%
\url{https://doi.org/10.1145/1031171.1031285}
\showDOI{\tempurl}


\bibitem[Kotu and Deshpande(2019)]%
        {Kotu2019}
\bibfield{author}{\bibinfo{person}{Vijay Kotu} {and} \bibinfo{person}{Bala Deshpande}.} \bibinfo{year}{2019}\natexlab{}.
\newblock \bibinfo{booktitle}{\emph{{Data Science Concepts and Practice}} (\bibinfo{edition}{second edi} ed.)}.
\newblock \bibinfo{publisher}{Morgan Kaufmann}.
\newblock
\showISBNx{978-0-12-814761-0}
\urldef\tempurl%
\url{https://doi.org/10.1016/C2017-0-02113-4}
\showDOI{\tempurl}


\bibitem[Kukkar and Mohana(2018)]%
        {Kukkar2018}
\bibfield{author}{\bibinfo{person}{Ashima Kukkar} {and} \bibinfo{person}{Rajni Mohana}.} \bibinfo{year}{2018}\natexlab{}.
\newblock \showarticletitle{A Supervised Bug Report Classification with Incorporate and Textual field Knowledge}.
\newblock \bibinfo{journal}{\emph{Procedia Computer Science}}  \bibinfo{volume}{132} (\bibinfo{year}{2018}), \bibinfo{pages}{352--361}.
\newblock
\showISSN{18770509}
\urldef\tempurl%
\url{https://doi.org/10.1016/j.procs.2018.05.194}
\showDOI{\tempurl}


\bibitem[Kukkar et~al\mbox{.}(2020)]%
        {Kukkar2020}
\bibfield{author}{\bibinfo{person}{Ashima Kukkar}, \bibinfo{person}{Rajni Mohana}, \bibinfo{person}{Yugal Kumar}, \bibinfo{person}{Anand Nayyar}, \bibinfo{person}{Muhammad Bilal}, {and} \bibinfo{person}{Kyung~Sup Kwak}.} \bibinfo{year}{2020}\natexlab{}.
\newblock \showarticletitle{{Duplicate Bug Report Detection and Classification System Based on Deep Learning Technique}}.
\newblock \bibinfo{journal}{\emph{IEEE Access}}  \bibinfo{volume}{8} (\bibinfo{year}{2020}), \bibinfo{pages}{200749--200763}.
\newblock
\showISSN{21693536}
\urldef\tempurl%
\url{https://doi.org/10.1109/ACCESS.2020.3033045}
\showDOI{\tempurl}


\bibitem[Kukkar et~al\mbox{.}(2019)]%
        {Kukkar2019}
\bibfield{author}{\bibinfo{person}{Ashima Kukkar}, \bibinfo{person}{Rajni Mohana}, \bibinfo{person}{Anand Nayyar}, \bibinfo{person}{Jeamin Kim}, \bibinfo{person}{Byeong~Gwon Kang}, {and} \bibinfo{person}{Naveen Chilamkurti}.} \bibinfo{year}{2019}\natexlab{}.
\newblock \showarticletitle{A novel deep-learning-based bug severity classification technique using convolutional neural networks and random forest with boosting}.
\newblock \bibinfo{journal}{\emph{Sensors (Switzerland)}}  \bibinfo{volume}{19} (\bibinfo{year}{2019}).
\newblock
Issue 13.
\showISSN{14248220}
\urldef\tempurl%
\url{https://doi.org/10.3390/s19132964}
\showDOI{\tempurl}
\newblock
\shownote{Read}.


\bibitem[{Kumar Nagwani} and Verma(2012)]%
        {KumarNagwani2012}
\bibfield{author}{\bibinfo{person}{Naresh {Kumar Nagwani}} {and} \bibinfo{person}{Shrish Verma}.} \bibinfo{year}{2012}\natexlab{}.
\newblock \showarticletitle{{CLUBAS: An Algorithm and Java Based Tool for Software Bug Classification Using Bug Attributes Similarities}}.
\newblock \bibinfo{journal}{\emph{Journal of Software Engineering and Applications}} \bibinfo{volume}{05}, \bibinfo{number}{06} (\bibinfo{year}{2012}), \bibinfo{pages}{436--447}.
\newblock
\showISSN{1945-3116}
\urldef\tempurl%
\url{https://doi.org/10.4236/jsea.2012.56050}
\showDOI{\tempurl}


\bibitem[Lee et~al\mbox{.}(2017)]%
        {Lee2017}
\bibfield{author}{\bibinfo{person}{Sun~Ro Lee}, \bibinfo{person}{Min~Jae Heo}, \bibinfo{person}{Chan~Gun Lee}, \bibinfo{person}{Milhan Kim}, {and} \bibinfo{person}{Gaeul Jeong}.} \bibinfo{year}{2017}\natexlab{}.
\newblock \showarticletitle{{Applying deep learning based automatic bug triager to industrial projects}}. In \bibinfo{booktitle}{\emph{Proceedings of the ACM SIGSOFT Symposium on the Foundations of Software Engineering}}, Vol.~\bibinfo{volume}{Part F130154}. \bibinfo{publisher}{Association for Computing Machinery}, \bibinfo{pages}{926--931}.
\newblock
\showISBNx{9781450351058}
\urldef\tempurl%
\url{https://doi.org/10.1145/3106237.3117776}
\showDOI{\tempurl}


\bibitem[Levene(1960)]%
        {Levene1960}
\bibfield{author}{\bibinfo{person}{H. Levene}.} \bibinfo{year}{1960}\natexlab{}.
\newblock \bibinfo{booktitle}{\emph{{Robust Tests for Equality of Variances}}}.
\newblock \bibinfo{publisher}{Stanford University Press}, \bibinfo{address}{Palo Alto}. 278--292 pages.
\newblock


\bibitem[Limsettho et~al\mbox{.}(2014)]%
        {Limsettho2014}
\bibfield{author}{\bibinfo{person}{Nachai Limsettho}, \bibinfo{person}{Hideaki Hata}, {and} \bibinfo{person}{Ken-ichi Matsumoto}.} \bibinfo{year}{2014}\natexlab{}.
\newblock \showarticletitle{{Comparing hierarchical dirichlet process with latent dirichlet allocation in bug report multiclass classification}}.
\newblock \bibinfo{journal}{\emph{15th IEEE/ACIS International Conference on Software Engineering, Artificial Intelligence, Networking and Parallel/Distributed Computing (SNPD)}} (\bibinfo{year}{2014}), \bibinfo{pages}{1--6}.
\newblock
\showISBNx{978-1-4799-5604-3}
\urldef\tempurl%
\url{https://doi.org/10.1109/SNPD.2014.6888695}
\showDOI{\tempurl}


\bibitem[Limsettho et~al\mbox{.}(2016)]%
        {Limsettho2016}
\bibfield{author}{\bibinfo{person}{Nachai Limsettho}, \bibinfo{person}{Hideaki Hata}, \bibinfo{person}{Akito Monden}, {and} \bibinfo{person}{Kenichi Matsumoto}.} \bibinfo{year}{2016}\natexlab{}.
\newblock \showarticletitle{{Unsupervised bug report categorization using clustering and labeling algorithm}}.
\newblock \bibinfo{journal}{\emph{International Journal of Software Engineering and Knowledge Engineering}} \bibinfo{volume}{26}, \bibinfo{number}{07} (\bibinfo{year}{2016}), \bibinfo{pages}{1027--1053}.
\newblock
\urldef\tempurl%
\url{https://doi.org/10.1142/S0218194016500352}
\showDOI{\tempurl}


\bibitem[Liu et~al\mbox{.}(2022)]%
        {Liu2022}
\bibfield{author}{\bibinfo{person}{Yong Liu}, \bibinfo{person}{Xuexin Qi}, \bibinfo{person}{Jiali Zhang}, \bibinfo{person}{Hui Li}, \bibinfo{person}{Xin Ge}, {and} \bibinfo{person}{Jun Ai}.} \bibinfo{year}{2022}\natexlab{}.
\newblock \showarticletitle{Automatic Bug Triaging via Deep Reinforcement Learning}.
\newblock \bibinfo{journal}{\emph{Applied Sciences (Switzerland)}}  \bibinfo{volume}{12} (\bibinfo{date}{4} \bibinfo{year}{2022}).
\newblock
Issue 7.
\showISSN{20763417}
\urldef\tempurl%
\url{https://doi.org/10.3390/app12073565}
\showDOI{\tempurl}
\newblock
\shownote{Read}.


\bibitem[Lopes et~al\mbox{.}(2020)]%
        {Lopes2020}
\bibfield{author}{\bibinfo{person}{Fábio Lopes}, \bibinfo{person}{João Agnelo}, \bibinfo{person}{César~A. Teixeira}, \bibinfo{person}{Nuno Laranjeiro}, {and} \bibinfo{person}{Jorge Bernardino}.} \bibinfo{year}{2020}\natexlab{}.
\newblock \showarticletitle{Automating orthogonal defect classification using machine learning algorithms}.
\newblock \bibinfo{journal}{\emph{Future Generation Computer Systems}}  \bibinfo{volume}{102} (\bibinfo{year}{2020}), \bibinfo{pages}{932--947}.
\newblock
\showISSN{0167739X}
\urldef\tempurl%
\url{https://doi.org/10.1016/j.future.2019.09.009}
\showDOI{\tempurl}


\bibitem[Manning et~al\mbox{.}(2008)]%
        {Manning2008}
\bibfield{author}{\bibinfo{person}{Christopher~D Manning}, \bibinfo{person}{Prabhakar Raghavan}, {and} \bibinfo{person}{Hinrich Sch{\"{u}}tze}.} \bibinfo{year}{2008}\natexlab{}.
\newblock \bibinfo{booktitle}{\emph{{Introduction to Information Retrieval}}}.
\newblock \bibinfo{publisher}{Cambridge University Press}.
\newblock
\urldef\tempurl%
\url{https://doi.org/10.1017/CBO9780511809071}
\showDOI{\tempurl}


\bibitem[Marwala(2018)]%
        {marwala2018}
\bibfield{author}{\bibinfo{person}{T Marwala}.} \bibinfo{year}{2018}\natexlab{}.
\newblock \bibinfo{booktitle}{\emph{{Handbook Of Machine Learning - Volume 1: Foundation Of Artificial Intelligence}}}.
\newblock \bibinfo{publisher}{World Scientific Publishing Company}.
\newblock
\showISBNx{9789813271241}
\urldef\tempurl%
\url{https://doi.org/10.1142/11013}
\showDOI{\tempurl}


\bibitem[Mian(2021)]%
        {Mian2021}
\bibfield{author}{\bibinfo{person}{Tariq~Saeed Mian}.} \bibinfo{year}{2021}\natexlab{}.
\newblock \showarticletitle{{Automation of Bug-Report Allocation to Developer using a Deep Learning Algorithm}}. In \bibinfo{booktitle}{\emph{2021 International Congress of Advanced Technology and Engineering (ICOTEN)}}. \bibinfo{pages}{1--7}.
\newblock
\urldef\tempurl%
\url{https://doi.org/10.1109/ICOTEN52080.2021.9493515}
\showDOI{\tempurl}


\bibitem[{Mingers Bsrcd}(1989)]%
        {MingersBsrcd1989}
\bibfield{author}{\bibinfo{person}{John {Mingers Bsrcd}}.} \bibinfo{year}{1989}\natexlab{}.
\newblock \showarticletitle{{An Empirical Comparison of Pruning Methods for Decision Tree Induction}}.
\newblock \bibinfo{journal}{\emph{Machine Learning}}  \bibinfo{volume}{4} (\bibinfo{year}{1989}), \bibinfo{pages}{227--243}.
\newblock
\urldef\tempurl%
\url{https://link.springer.com/content/pdf/10.1023%2FA%3A1022604100933.pdf}
\showURL{%
\tempurl}


\bibitem[Mueller and Massaron(2021)]%
        {mueller2021}
\bibfield{author}{\bibinfo{person}{J~P Mueller} {and} \bibinfo{person}{L Massaron}.} \bibinfo{year}{2021}\natexlab{}.
\newblock \bibinfo{booktitle}{\emph{{Machine Learning For Dummies}}}.
\newblock \bibinfo{publisher}{Wiley}.
\newblock
\showISBNx{9781119724018}
\urldef\tempurl%
\url{https://books.google.pt/books?id=B-USEAAAQBAJ}
\showURL{%
\tempurl}


\bibitem[Novakovi{\'{c}} et~al\mbox{.}(2011)]%
        {Novakovic2011}
\bibfield{author}{\bibinfo{person}{Jasmina Novakovi{\'{c}}}, \bibinfo{person}{Perica Strbac}, {and} \bibinfo{person}{Dusan Bulatovi{\'{c}}}.} \bibinfo{year}{2011}\natexlab{}.
\newblock \showarticletitle{{Toward optimal feature selection using ranking methods and classification algorithms}}.
\newblock \bibinfo{journal}{\emph{Yugoslav Journal of Operations Research}} \bibinfo{volume}{21}, \bibinfo{number}{1} (\bibinfo{year}{2011}), \bibinfo{pages}{119--135}.
\newblock
\showISSN{03540243}
\urldef\tempurl%
\url{https://doi.org/10.2298/YJOR1101119N}
\showDOI{\tempurl}


\bibitem[Panda and Nagwani(2019)]%
        {Panda2019}
\bibfield{author}{\bibinfo{person}{Rama~Ranjan Panda} {and} \bibinfo{person}{Naresh~Kumar Nagwani}.} \bibinfo{year}{2019}\natexlab{}.
\newblock \showarticletitle{Software Bug Categorization Technique Based on Fuzzy Similarity}.
\newblock \bibinfo{journal}{\emph{Proceedings of the 2019 IEEE 9th International Conference on Advanced Computing, IACC 2019}} (\bibinfo{year}{2019}), \bibinfo{pages}{1--6}.
\newblock
\showISBNx{9781728143927}
\urldef\tempurl%
\url{https://doi.org/10.1109/IACC48062.2019.8971599}
\showDOI{\tempurl}


\bibitem[Pandey et~al\mbox{.}(2017)]%
        {Pandey2017}
\bibfield{author}{\bibinfo{person}{Nitish Pandey}, \bibinfo{person}{Debarshi~Kumar Sanyal}, \bibinfo{person}{Abir Hudait}, {and} \bibinfo{person}{Amitava Sen}.} \bibinfo{year}{2017}\natexlab{}.
\newblock \showarticletitle{Automated classification of software issue reports using machine learning techniques: an empirical study}.
\newblock \bibinfo{journal}{\emph{Innovations in Systems and Software Engineering}}  \bibinfo{volume}{13} (\bibinfo{year}{2017}), \bibinfo{pages}{279--297}.
\newblock
Issue 4.
\showISBNx{1133401702941}
\showISSN{16145054}
\urldef\tempurl%
\url{https://doi.org/10.1007/s11334-017-0294-1}
\showDOI{\tempurl}


\bibitem[Pingclasai et~al\mbox{.}(2013)]%
        {Pingclasai2013}
\bibfield{author}{\bibinfo{person}{Natthakul Pingclasai}, \bibinfo{person}{Hideaki Hata}, {and} \bibinfo{person}{Ken~Ichi Matsumoto}.} \bibinfo{year}{2013}\natexlab{}.
\newblock \showarticletitle{{Classifying bug reports to bugs and other requests using topic modeling}}.
\newblock \bibinfo{journal}{\emph{Proceedings - Asia-Pacific Software Engineering Conference, APSEC}}  \bibinfo{volume}{2} (\bibinfo{year}{2013}), \bibinfo{pages}{13--18}.
\newblock
\showISBNx{9781479921430}
\showISSN{15301362}
\urldef\tempurl%
\url{https://doi.org/10.1109/APSEC.2013.105}
\showDOI{\tempurl}


\bibitem[Pushpalatha and Mrunalini(2019)]%
        {Pushpalatha2019}
\bibfield{author}{\bibinfo{person}{M.~N. Pushpalatha} {and} \bibinfo{person}{M. Mrunalini}.} \bibinfo{year}{2019}\natexlab{}.
\newblock \showarticletitle{Predicting the severity of open source bug reports using unsupervised and supervised techniques}.
\newblock \bibinfo{journal}{\emph{International Journal of Open Source Software and Processes}}  \bibinfo{volume}{10} (\bibinfo{year}{2019}), \bibinfo{pages}{1--15}.
\newblock
Issue 1.
\showISSN{19423934}
\urldef\tempurl%
\url{https://doi.org/10.4018/IJOSSP.2019010101}
\showDOI{\tempurl}


\bibitem[Qin and Sun(2018)]%
        {Qin2018}
\bibfield{author}{\bibinfo{person}{Hanmin Qin} {and} \bibinfo{person}{Xin Sun}.} \bibinfo{year}{2018}\natexlab{}.
\newblock \showarticletitle{{Classifying bug reports into bugs and non-bugs using LSTM}}.
\newblock \bibinfo{journal}{\emph{ACM International Conference Proceeding Series}} (\bibinfo{year}{2018}), \bibinfo{pages}{16--19}.
\newblock
\showISBNx{9781450365901}
\urldef\tempurl%
\url{https://doi.org/10.1145/3275219.3275239}
\showDOI{\tempurl}


\bibitem[Sammut and Webb(2011)]%
        {Sammut2011}
\bibfield{author}{\bibinfo{person}{Claude Sammut} {and} \bibinfo{person}{Geoffrey~I Webb}.} \bibinfo{year}{2011}\natexlab{}.
\newblock \bibinfo{booktitle}{\emph{Encyclopedia of Machine Learning} (\bibinfo{edition}{1st} ed.)}.
\newblock \bibinfo{publisher}{Springer Publishing Company, Incorporated}.
\newblock
\showISBNx{0387307680}


\bibitem[Shapiro and Wilk(1965)]%
        {Shapiro1965}
\bibfield{author}{\bibinfo{person}{S~S Shapiro} {and} \bibinfo{person}{M~B Wilk}.} \bibinfo{year}{1965}\natexlab{}.
\newblock \showarticletitle{{An Analysis of Variance Test for Normality (Complete Samples)}}.
\newblock \bibinfo{journal}{\emph{Biometrika}} \bibinfo{volume}{52}, \bibinfo{number}{3/4} (\bibinfo{year}{1965}), \bibinfo{pages}{591--611}.
\newblock
\showISSN{00063444}
\urldef\tempurl%
\url{http://www.jstor.org/stable/2333709}
\showURL{%
\tempurl}


\bibitem[Shatnawi and Alazzam(2022)]%
        {Shatnawi2022}
\bibfield{author}{\bibinfo{person}{Mohammed~Q. Shatnawi} {and} \bibinfo{person}{Batool Alazzam}.} \bibinfo{year}{2022}\natexlab{}.
\newblock \showarticletitle{{An Assessment of Essessment Eclipse Bugs' Priority and Severity Prediction Using Machine Learning}}.
\newblock \bibinfo{journal}{\emph{International Journal of Communication Networks and Information Security}} \bibinfo{volume}{14}, \bibinfo{number}{1} (\bibinfo{year}{2022}), \bibinfo{pages}{62--69}.
\newblock
\showISSN{2073607X}
\urldef\tempurl%
\url{https://doi.org/10.54039/ijcnis.v14i1.5266}
\showDOI{\tempurl}


\bibitem[Sohrawardi et~al\mbox{.}(2014)]%
        {Sohrawardi2014}
\bibfield{author}{\bibinfo{person}{Saniat~Javid Sohrawardi}, \bibinfo{person}{Iftekhar Azam}, {and} \bibinfo{person}{Shazzad Hosain}.} \bibinfo{year}{2014}\natexlab{}.
\newblock \showarticletitle{A comparative study of text classification algorithms on user submitted bug reports}.
\newblock \bibinfo{journal}{\emph{2014 9th International Conference on Digital Information Management, ICDIM 2014}} (\bibinfo{year}{2014}), \bibinfo{pages}{242--247}.
\newblock
Issue May.
\showISBNx{9781479954209}
\urldef\tempurl%
\url{https://doi.org/10.1109/ICDIM.2014.6991434}
\showDOI{\tempurl}


\bibitem[Srinivasa-Desikan(2018)]%
        {Srinivasa2018}
\bibfield{author}{\bibinfo{person}{Bhargav. Srinivasa-Desikan}.} \bibinfo{year}{2018}\natexlab{}.
\newblock \bibinfo{booktitle}{\emph{Natural Language Processing and Computational Linguistics : a Practical Guide to Text Analysis with Python, Gensim, SpaCy, and Keras.}}
\newblock \bibinfo{publisher}{Packt Publishing Ltd}. 298 pages.
\newblock
\showISBNx{9781788838535}


\bibitem[Syarif et~al\mbox{.}(2016)]%
        {Syarif2016}
\bibfield{author}{\bibinfo{person}{Iwan Syarif}, \bibinfo{person}{Adam Prugel-Bennett}, {and} \bibinfo{person}{Gary Wills}.} \bibinfo{year}{2016}\natexlab{}.
\newblock \showarticletitle{{SVM Parameter Optimization using Grid Search and Genetic Algorithm to Improve Classification Performance}}.
\newblock \bibinfo{journal}{\emph{TELKOMNIKA (Telecommunication Computing Electronics and Control)}} \bibinfo{volume}{14}, \bibinfo{number}{4} (\bibinfo{year}{2016}), \bibinfo{pages}{1502}.
\newblock
\showISSN{1693-6930}
\urldef\tempurl%
\url{https://doi.org/10.12928/telkomnika.v14i4.3956}
\showDOI{\tempurl}


\bibitem[Terdchanakul et~al\mbox{.}(2017)]%
        {Terdchanakul2017}
\bibfield{author}{\bibinfo{person}{Pannavat Terdchanakul}, \bibinfo{person}{Hideaki Hata}, \bibinfo{person}{Passakorn Phannachitta}, {and} \bibinfo{person}{Kenichi Matsumoto}.} \bibinfo{year}{2017}\natexlab{}.
\newblock \showarticletitle{Bug or not? Bug Report classification using N-gram IDF}.
\newblock \bibinfo{journal}{\emph{Proceedings - 2017 IEEE International Conference on Software Maintenance and Evolution, ICSME 2017}} (\bibinfo{year}{2017}), \bibinfo{pages}{534--538}.
\newblock
\showISBNx{9781538609927}
\urldef\tempurl%
\url{https://doi.org/10.1109/ICSME.2017.14}
\showDOI{\tempurl}


\bibitem[Thanaki(2017)]%
        {thanaki2017}
\bibfield{author}{\bibinfo{person}{Jalaj Thanaki}.} \bibinfo{year}{2017}\natexlab{}.
\newblock \bibinfo{booktitle}{\emph{Python Natural Language Processing}}.
\newblock
\showISBNx{9781787121423}
\urldef\tempurl%
\url{www.packtpub.com}
\showURL{%
\tempurl}


\bibitem[Theobald(2024)]%
        {theobald2024}
\bibfield{author}{\bibinfo{person}{O Theobald}.} \bibinfo{year}{2024}\natexlab{}.
\newblock \bibinfo{booktitle}{\emph{{Machine Learning with Python: Unlocking AI Potential with Python and Machine Learning}}}.
\newblock \bibinfo{publisher}{Packt Publishing}.
\newblock
\showISBNx{9781835462072}
\urldef\tempurl%
\url{https://books.google.pt/books?id=6gv6EAAAQBAJ}
\showURL{%
\tempurl}


\bibitem[Uysal and Gunal(2014)]%
        {Uysal2014}
\bibfield{author}{\bibinfo{person}{Alper~Kursat Uysal} {and} \bibinfo{person}{Serkan Gunal}.} \bibinfo{year}{2014}\natexlab{}.
\newblock \showarticletitle{{The impact of preprocessing on text classification}}.
\newblock \bibinfo{journal}{\emph{Information Processing \& Management}} \bibinfo{volume}{50}, \bibinfo{number}{1} (\bibinfo{year}{2014}), \bibinfo{pages}{104--112}.
\newblock
\showISSN{0306-4573}
\urldef\tempurl%
\url{https://doi.org/10.1016/j.ipm.2013.08.006}
\showDOI{\tempurl}


\bibitem[Vajjala et~al\mbox{.}(2020)]%
        {vajjala2020}
\bibfield{author}{\bibinfo{person}{S Vajjala}, \bibinfo{person}{B Majumder}, \bibinfo{person}{A Gupta}, {and} \bibinfo{person}{H Surana}.} \bibinfo{year}{2020}\natexlab{}.
\newblock \bibinfo{booktitle}{\emph{Practical Natural Language Processing: A Comprehensive Guide to Building Real-World NLP Systems}}.
\newblock \bibinfo{publisher}{O'Reilly Media}.
\newblock
\showISBNx{9781492054023}


\bibitem[Verleysen and Fran{\c{c}}ois(2005)]%
        {Verleysen2005}
\bibfield{author}{\bibinfo{person}{Michel Verleysen} {and} \bibinfo{person}{Damien Fran{\c{c}}ois}.} \bibinfo{year}{2005}\natexlab{}.
\newblock \showarticletitle{{The Curse of Dimensionality in Data Mining and Time Series Prediction}}. In \bibinfo{booktitle}{\emph{Computational Intelligence and Bioinspired Systems}}, \bibfield{editor}{\bibinfo{person}{Joan Cabestany}, \bibinfo{person}{Alberto Prieto}, {and} \bibinfo{person}{Francisco Sandoval}} (Eds.). \bibinfo{publisher}{Springer Berlin Heidelberg}, \bibinfo{address}{Berlin, Heidelberg}, \bibinfo{pages}{758--770}.
\newblock
\showISBNx{978-3-540-32106-4}


\bibitem[Wang et~al\mbox{.}(2016)]%
        {Wang2016a}
\bibfield{author}{\bibinfo{person}{Junjie Wang}, \bibinfo{person}{Song Wang}, \bibinfo{person}{Qiang Cui}, {and} \bibinfo{person}{Qing Wang}.} \bibinfo{year}{2016}\natexlab{}.
\newblock \showarticletitle{{Local-based active classification of test report to assist crowdsourced testing}}.
\newblock \bibinfo{journal}{\emph{ASE 2016 - Proceedings of the 31st IEEE/ACM International Conference on Automated Software Engineering}} (\bibinfo{year}{2016}), \bibinfo{pages}{190--201}.
\newblock
\showISBNx{9781450338455}
\urldef\tempurl%
\url{https://doi.org/10.1145/2970276.2970300}
\showDOI{\tempurl}


\bibitem[Wilcoxon et~al\mbox{.}(1963)]%
        {Wilcoxon1963}
\bibfield{author}{\bibinfo{person}{F Wilcoxon}, \bibinfo{person}{S~K Katti}, {and} \bibinfo{person}{R~A Wilcox}.} \bibinfo{year}{1963}\natexlab{}.
\newblock \bibinfo{booktitle}{\emph{{Critical Values and Probability Levels for the Wilcoxon Rank Sum Test and the Wilcoxon Signed Rank Test}}}.
\newblock \bibinfo{publisher}{American Cyanamid}.
\newblock
\urldef\tempurl%
\url{https://books.google.com.br/books?id=GS3aGAAACAAJ}
\showURL{%
\tempurl}


\bibitem[Yap et~al\mbox{.}(2014)]%
        {Yap2014}
\bibfield{author}{\bibinfo{person}{Bee~Wah Yap}, \bibinfo{person}{Khatijahhusna~Abd Rani}, \bibinfo{person}{Hezlin~Aryani {Abd Rahman}}, \bibinfo{person}{Simon Fong}, \bibinfo{person}{Zuraida Khairudin}, {and} \bibinfo{person}{Nik~Nairan Abdullah}.} \bibinfo{year}{2014}\natexlab{}.
\newblock \showarticletitle{{An application of oversampling, undersampling, bagging and boosting in handling imbalanced datasets}}.
\newblock \bibinfo{journal}{\emph{Lecture Notes in Electrical Engineering}}  \bibinfo{volume}{285 LNEE} (\bibinfo{year}{2014}), \bibinfo{pages}{13--22}.
\newblock
\showISBNx{9789814585170}
\showISSN{18761119}
\urldef\tempurl%
\url{https://doi.org/10.1007/978-981-4585-18-7_2}
\showDOI{\tempurl}


\bibitem[Zhang et~al\mbox{.}(2020)]%
        {Zhang2020}
\bibfield{author}{\bibinfo{person}{Tian~Lun Zhang}, \bibinfo{person}{Rong Chen}, \bibinfo{person}{Xi Yang}, {and} \bibinfo{person}{Hong~Yu Zhu}.} \bibinfo{year}{2020}\natexlab{}.
\newblock \showarticletitle{An uncertainty based incremental learning for identifying the severity of bug report}.
\newblock \bibinfo{journal}{\emph{International Journal of Machine Learning and Cybernetics}}  \bibinfo{volume}{11} (\bibinfo{year}{2020}), \bibinfo{pages}{123--136}.
\newblock
Issue 1.
\showISBNx{1304201900961}
\showISSN{1868808X}
\urldef\tempurl%
\url{https://doi.org/10.1007/s13042-019-00961-2}
\showDOI{\tempurl}
\newblock
\shownote{Ok.<br/>14 pages... leave this one to the end}.


\bibitem[Zhou et~al\mbox{.}(2016)]%
        {Zhou2016}
\bibfield{author}{\bibinfo{person}{Yu Zhou}, \bibinfo{person}{Yanxiang Tong}, \bibinfo{person}{Ruihang Gu}, {and} \bibinfo{person}{Harald Gall}.} \bibinfo{year}{2016}\natexlab{}.
\newblock \showarticletitle{{Combining text mining and data mining for bug report classification}}.
\newblock \bibinfo{journal}{\emph{Journal of Software: Evolution and Process}} \bibinfo{volume}{28}, \bibinfo{number}{3} (\bibinfo{year}{2016}), \bibinfo{pages}{150--176}.
\newblock
\urldef\tempurl%
\url{https://doi.org/10.1002/smr.1770}
\showDOI{\tempurl}


\bibitem[Zhu et~al\mbox{.}(2019)]%
        {Zhu2019}
\bibfield{author}{\bibinfo{person}{Yuxiang Zhu}, \bibinfo{person}{Minxue Pan}, \bibinfo{person}{Yu Pei}, {and} \bibinfo{person}{Tian Zhang}.} \bibinfo{year}{2019}\natexlab{}.
\newblock \showarticletitle{{A Bug or a Suggestion? An Automatic Way to Label Issues}}.
\newblock \bibinfo{journal}{\emph{arXiv preprint arXiv:1909.00934}} (\bibinfo{year}{2019}).
\newblock
\showeprint[arxiv]{1909.00934}
\urldef\tempurl%
\url{http://arxiv.org/abs/1909.00934}
\showURL{%
\tempurl}


\bibitem[Zimmermann et~al\mbox{.}(2010)]%
        {Zimmermann2010}
\bibfield{author}{\bibinfo{person}{Thomas Zimmermann}, \bibinfo{person}{Rahul Premraj}, \bibinfo{person}{Nicolas Bettenburg}, \bibinfo{person}{Sascha Just}, \bibinfo{person}{Adrian Schröter}, {and} \bibinfo{person}{Cathrin Weiss}.} \bibinfo{year}{2010}\natexlab{}.
\newblock \showarticletitle{What makes a good bug report?}
\newblock \bibinfo{journal}{\emph{IEEE Transactions on Software Engineering}}  \bibinfo{volume}{36} (\bibinfo{year}{2010}), \bibinfo{pages}{618--643}.
\newblock
Issue 5.
\showISSN{00985589}
\urldef\tempurl%
\url{https://doi.org/10.1109/TSE.2010.63}
\showDOI{\tempurl}


\end{thebibliography}

\end{document}